\documentclass[11pt,twoside,a4paper,cmspaper,final,collab]{cms-tdr}
\def\svnVersion{52d7fbc-D}\def\svnDate{2026/10/23}\def\cmsCernNoTag{CERN-EP-2026-260}\def\cmsCernDate{\today}\def\cmsMessage{Submitted to the Journal of High Energy Physics}
\begin{document}\cmsNoteHeader{EXO-24-010}

\newcommand{\zjets}{{{\PZ}+jets}\xspace}
\newcommand{\wjets}{{{\PW}+jets}\xspace}
\newcommand{\znnjets}{\ensuremath{{\cPZ}{(\to\PGn\PAGn)\text{+jets}}}\xspace}
\newcommand{\wlnjets}{\ensuremath{{\PW}{(\to\ell\PGn)\text{+jets}}}\xspace}
\newcommand{\ttjets}{{\ttbar}+jets\xspace}
\newcommand{\MADGRAPHFULL}{\MGvATNLO\,2.6.5\xspace}
\newcommand{\PYTHIAFULL}{\PYTHIA\,8.240\xspace}
\newcommand{\FEYNRULES}{\textsc{FeynRules}\xspace}
\newcommand{\PNET}{\textsc{ParticleNet}\xspace}
\newcommand{\mt}{\ensuremath{m_{\mathrm{T}}}\xspace}
\renewcommand{\mT}{\mt}
\newcommand{\ST}{\ensuremath{S_{\mathrm{T}}}\xspace}
\newcommand{\widejet}{\ensuremath{\mathrm{J}}\xspace}
\newcommand{\narrowjet}{\ensuremath{\mathrm{j}}\xspace}
\newcommand{\dR}{\ensuremath{\Delta R}\xspace}
\newcommand{\Deta}{\ensuremath{\Delta\eta}\xspace}
\newcommand{\Dphi}{\ensuremath{\Delta\phi}\xspace}
\newcommand{\mindphi}{\ensuremath{\Dphi_{\text{min}}}\xspace}
\newcommand{\mindphifull}{\ensuremath{\mindphi(\widejet,\ptvecmiss)}\xspace}
\newcommand{\Imini}{\ensuremath{I_{\text{mini}}}\xspace}
\newcommand{\ptsub}[1]{\ensuremath{p_{\text{T},#1}}\xspace}
\newcommand{\ptD}{\ensuremath{D_{\pt}}\xspace}
\newcommand{\nsvj}{\ensuremath{n_{\text{SVJ}}}\xspace}
\newcommand{\nsvjwnae}{\ensuremath{n_{\text{SVJ}}^{\text{WNAE}}}\xspace}
\newcommand{\nsvjpn}{\ensuremath{n_{\text{SVJ}}^{\text{PN}}}\xspace}
\newcommand{\PPhi}{\ensuremath{\Phi}\xspace}
\newcommand{\mPhi}{\ensuremath{m_{\PPhi}}\xspace}
\newcommand{\sigmaPhi}{\ensuremath{\sigma_{\PPhi}}\xspace}
\newcommand{\abseta}{\ensuremath{\abs{\eta}}\xspace}
\newcommand{\dRmax}{\ensuremath{\dR_{\text{max}}}\xspace}
\newcommand{\Pchhad}{\ensuremath{\mathrm{h}^{\pm}}\xspace}
\newcommand{\Pnhad}{\ensuremath{\mathrm{h}^{0}}\xspace}
\newcommand{\mDark}{\ensuremath{m_{\text{dark}}}\xspace}
\newcommand{\mDarkMax}{\ensuremath{m_{\text{dark}}^{\text{max}}}\xspace}
\newcommand{\lamDark}{\ensuremath{\Lambda_{\text{dark}}}\xspace}
\newcommand{\rinv}{\ensuremath{r_{\text{inv}}}\xspace}
\newcommand{\yukawa}{\ensuremath{\lambda}\xspace}
\newcommand{\Pqdark}{\PGc}
\newcommand{\mqdark}{\ensuremath{m_{\Pqdark}}\xspace}
\newcommand{\Paqdark}{\HepAntiParticle{\PGc}{}{}\xspace}
\newcommand{\PqdarkO}{\HepParticle{\PGc}{}{1}\xspace}
\newcommand{\PqdarkT}{\HepParticle{\PGc}{}{2}\xspace}
\newcommand{\Pgdark}{\ensuremath{\Pg_{\text{dark}}}\xspace}
\newcommand{\Ppidark}{\ensuremath{\PGp_{\text{dark}}}\xspace}
\newcommand{\PpidarkDM}{\ensuremath{\Ppidark^{\text{DM}}}\xspace}
\newcommand{\Prhodark}{\ensuremath{\PGr_{\text{dark}}}\xspace}
\newcommand{\PrhodarkDM}{\ensuremath{\Prhodark^{\text{DM}}}\xspace}
\newcommand{\Ncdark}{\ensuremath{N_{c}^{\text{dark}}}\xspace}
\newcommand{\Nfdark}{\ensuremath{N_{f}^{\text{dark}}}\xspace}
\newcommand{\mq}{\ensuremath{m_{\PQq}}\xspace}
\newcommand{\Nstable}{\ensuremath{N_{\text{stable}}}\xspace}
\newcommand{\Nunstable}{\ensuremath{N_{\text{unstable}}}\xspace}
\newcommand{\yijab}{\ensuremath{\yukawa_{ijab}}\xspace}
\newcommand{\NApred}{\ensuremath{N_{\text{A}}^{\text{pred}}}\xspace}
\newcommand{\NAa}{\ensuremath{N_\text{A}}\xspace}
\newcommand{\NB}{\ensuremath{N_\text{B}}\xspace}
\newcommand{\NC}{\ensuremath{N_\text{C}}\xspace}
\newcommand{\ND}{\ensuremath{N_\text{D}}\xspace}
\newcommand{\supervised}{supervised tagger strategy\xspace}
\newcommand{\unsupervised}{unsupervised tagger strategy\xspace}
\newcommand{\strategies}{supervised and unsupervised tagger strategies\xspace}
\newcommand{\supervisedA}{supervised tagger analysis\xspace}
\newcommand{\unsupervisedA}{unsupervised tagger analysis\xspace}

\newlength\cmsTabSkip\setlength{\cmsTabSkip}{1ex}
\providecommand{\cmsTable}[1]{\resizebox{\textwidth}{!}{#1}}

\cmsNoteHeader{EXO-24-010}

\title{Search for nonresonant production of strongly coupled dark matter in proton-proton collisions at \texorpdfstring{$\sqrt{s}=13\TeV$}{sqrt(s) = 13 TeV}}

\date{\today}

\abstract{
A search is presented for semivisible jets produced in nonresonant processes, using proton-proton collision data at $\sqrt{s}=13\TeV$, collected with the CMS detector at the CERN LHC and corresponding to an integrated luminosity of 138\fbinv. These jets, containing both visible and invisible particles, would indicate the presence of dark matter arising from a strongly coupled hidden sector. The signal model includes a bifundamental scalar that couples simultaneously to standard model quarks and dark-sector quarks, the latter forming semivisible jets. The final state therefore includes missing transverse momentum aligned with a jet, rather than recoiling against standard model particles as typically required by conventional collider dark-matter searches. Two separate machine-learning strategies are employed to identify semivisible jets: a supervised graph neural network and an unsupervised autoencoder. No significant excess corresponding to dark-sector signal hypotheses is observed. Upper limits at 95\% confidence level on the signal production cross section are determined for each strategy and interpreted as exclusions in the signal model parameter space. The supervised search excludes mediator masses between 0.5 and 2.4\TeV, depending on the other signal model parameters. For signal models with semivisible jets containing higher fractions of stable dark-matter particles, the unsupervised search demonstrates increased sensitivity with 5 (11)\% higher expected (observed) mediator mass exclusions than the supervised search.
}

\hypersetup{%
pdfauthor={CMS Collaboration}, %
pdftitle={Search for nonresonant production of strongly coupled dark matter in proton-proton collisions at sqrt(s) = 13 TeV},%
pdfsubject={CMS},%
pdfkeywords={CMS, dark showers, semivisible jets, dark matter, beyond standard model}}

\maketitle 

\section{Introduction}\label{sec:intro}

For several decades, precision measurements at particle colliders have provided insights into elementary particles, such as electroweak bosons, the top quark, and, more recently, the Higgs boson.
These experimental results have been found to agree with the theoretical predictions of the standard model (SM) of particle physics.
However, a wealth of astronomical observations points to the existence of dark matter (DM), unexplained by the SM.
Observations such as deviations in galaxy rotation curves~\cite{Rubin:1980zd,Persic:1995ru}, gravitational lensing in galaxy cluster collisions~\cite{Clowe:2006eq} and in large-scale structures~\cite{Chang:2017kmv}, and the cosmic microwave background power spectrum~\cite{Planck:2018vyg} strongly indicate the presence of an unknown form of matter.

Numerous extensions of the SM that incorporate DM and remain consistent with astrophysical observations predict interactions other than gravitational between DM and SM particles.
Collider-based investigations have predominantly focused on weakly interacting massive particles (WIMPs), which are among the most theoretically well-motivated DM candidates.
These particles potentially lie within a mass range suitable for production at colliders~\cite{Jungman:1995df} and typically produce distinctive experimental signatures, where missing transverse momentum \ptmiss is observed recoiling against visible SM particles.
The CMS and ATLAS Collaborations have searched for DM by analyzing various signatures accompanying the \ptmiss, such as jets or leptons~\cite{CMS:2018ffd,CMS:2020ulv,CMS:2021far, Sirunyan:2018owy,CMS:2023dof,ATLAS:2024rlu,Aaboud:2018sfi,ATLAS:2019wdu,ATLAS:2021hza,ATLAS:2020xzu,ATLAS:2021kxv,ATLAS:2024kpy}, Higgs bosons~\cite{Sirunyan:2018gdw,CMS:2019ykj}, or top quarks~\cite{Sirunyan:2019gfm,ATLAS:2022znu,CMS:2025ncs}, as summarized in Refs.~\cite{CMS:2024zqs,ATLAS:2023rvb,ATLAS:2024qmx}.
Despite extensive efforts, no evidence for WIMP DM has been observed thus far.

As an alternative, more complex DM models, such as hidden valleys (HVs)~\cite{Strassler:2006im}, predict the existence of dark sectors (DSs) containing beyond-the-SM particles that interact with the SM sector via specific mediators.
In particular, strongly coupled DSs governed by a confining interaction resulting from an $SU(N)$ gauge symmetry, similar to the SM quantum chromodynamics (QCD) and therefore called ``dark QCD'', have only begun to be probed in recent years.
Such scenarios predict unique signatures arising from dark showers~\cite{Cohen:2017pzm}.

\begin{figure}[htb!]
\centering
\includegraphics[height=0.18\linewidth]{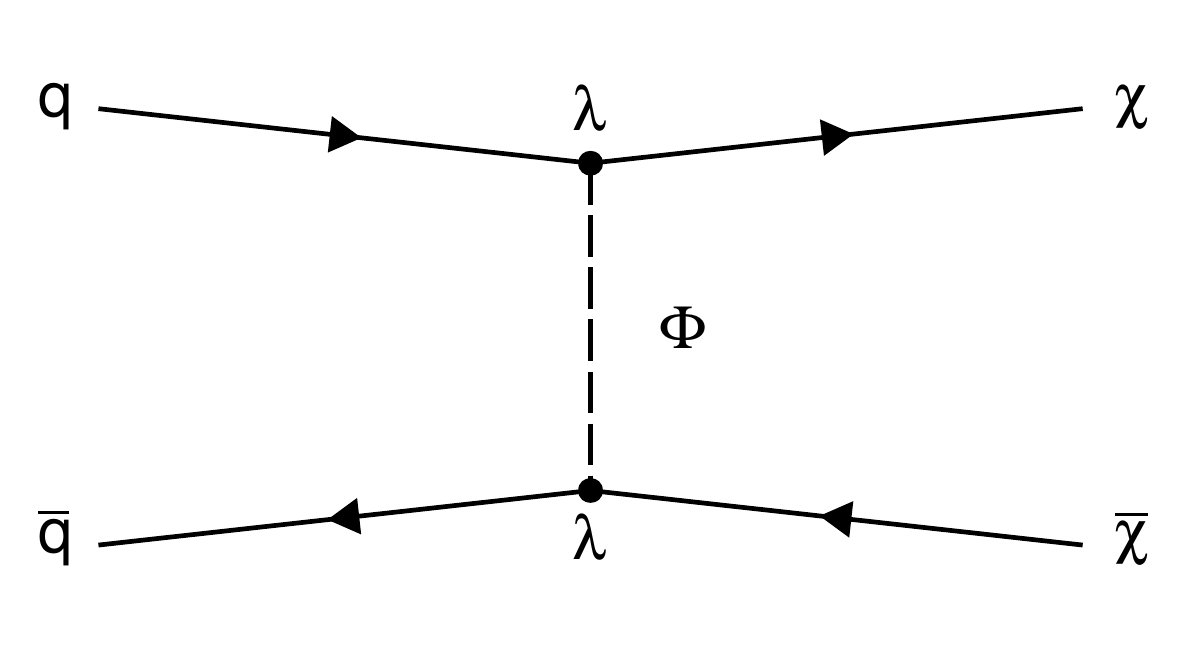}
\includegraphics[height=0.18\linewidth]{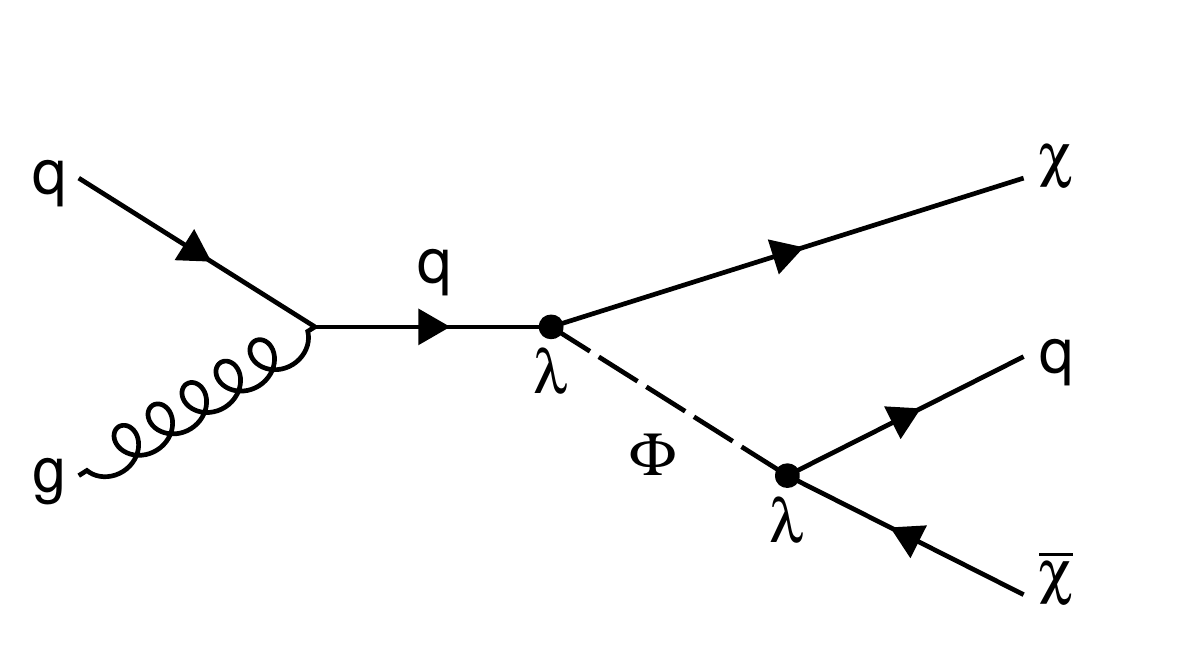}
\includegraphics[height=0.18\linewidth]{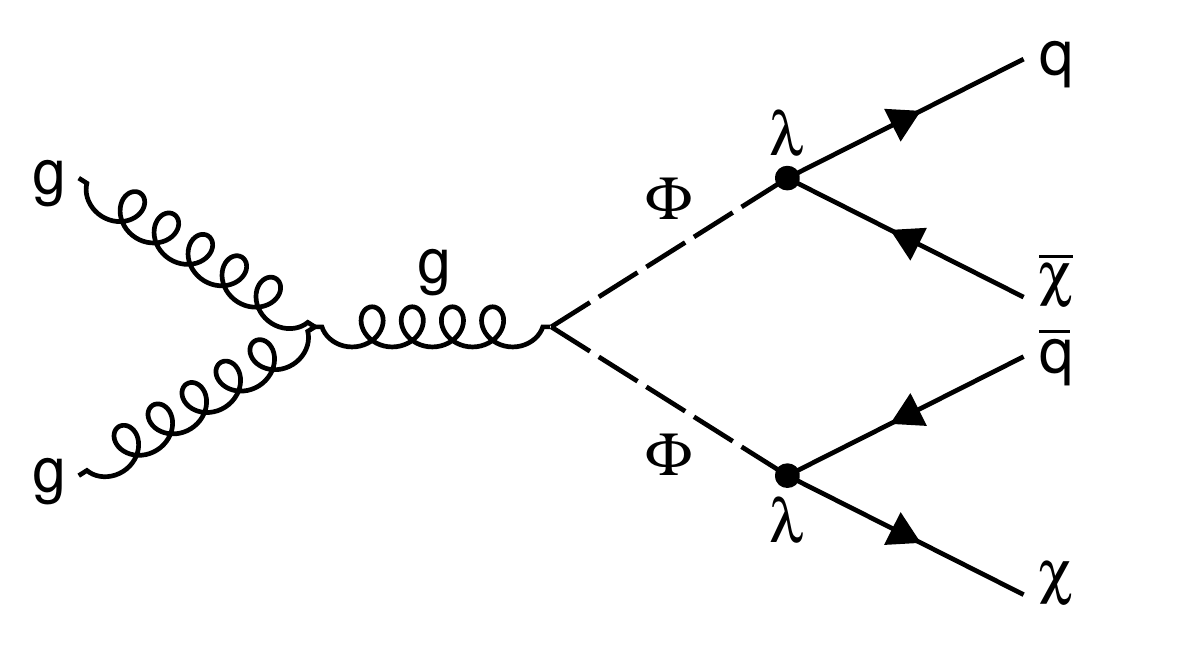}
\caption{Representative Feynman diagrams of leading-order production of dark quarks \Pqdark through a bifundamental mediator \PPhi. Left: $t$-channel production of dark quarks. Middle: associated production of \PPhi with a dark quark. Right: \PPhi pair production.}
\label{fig:t-chan-diagram}
\end{figure}

This paper investigates a class of models proposed in Ref.~\cite{Cohen:2017pzm} that predict a DS containing several flavors of dark quarks that shower and hadronize in the DS, forming stable and unstable bound states, called dark hadrons.
Stable dark hadrons---DM candidates---escape the detector without interacting, while unstable ones decay promptly to SM quarks, which further shower and hadronize in the visible sector.
As a result, the jets produced in these events consist of both visible and invisible components, leading to a distinctive collider signature referred to as ``semivisible jets'' (SVJs).
The signal models under study in this analysis focus on a nonresonant production mechanism wherein a massive scalar bifundamental mediator \PPhi provides the link between the dark and visible sectors, as illustrated in Fig.~\ref{fig:t-chan-diagram}.
The mediator \PPhi is in the fundamental representation under both the SM QCD and the dark QCD gauge groups.
The coupling to dark and visible sectors is Yukawa-like and is defined by:
\begin{equation}
\mathcal{L}_{\PPhi} \supset \sum_{i,j,a,b}\yijab\Paqdark_{a}\PPhi^{*}_{bi}\PQq_{j},
\end{equation} 
where $i, j$ are the SM flavor indices; $a,b$ are the dark-flavor indices; $\Paqdark_{a}$, $\PQq_{j}$, and $\PPhi_{bi}$ denote the dark-antiquark, SM quark, and bifundamental scalar fields, respectively; and \yijab is the Yukawa coupling.
This scenario allows a variety of production modes, including the $t$-channel production of a pair of dark quarks, the associated production of a dark quark and \PPhi, and the \PPhi pair production.
Depending on \yijab and the mediator mass \mPhi, the contribution of each process to the total cross section varies.
Representative Feynman diagrams of the three production processes are illustrated in Fig.~\ref{fig:t-chan-diagram}.
Further details about the signal model and the simulated samples used in the analysis are provided in Section~\ref{sec:DataSim}.

The stable and unstable dark hadrons are produced together in collimated sprays.
Therefore, a typical final-state signature includes jets from the visible decay products of unstable dark hadrons with nonnegligible \ptmiss from the invisible dark hadrons, aligned with the jets.
If the dark-hadron mass scale is higher than the SM light-quark mass scale, the visible decay products tend to be produced with significant momentum in the dark-hadron rest frame, such that they spread out to produce wider jets in the lab frame than typical SM processes.
This effect, compounded by the fragmentation and showering processes that occur in both the DS and the SM sector, results in observable differences in jet substructure between SVJs and SM jets.
Section~\ref{sec:EvtReco} describes the reconstruction of these jets and other objects, along with the selection of relevant events.

The CMS and ATLAS Collaborations have searched for several other types of dark-shower signatures, including ``emerging jets'', which arise from long-lived dark hadrons~\cite{Schwaller:2015gea, Sirunyan:2018njd, CMS:2024gxp, ATLAS:2025bsz, ATLAS:2025lfx}, and ``soft unclustered energy patterns'', which are characterized by many low-momentum, isotropically distributed particles~\cite{Strassler:2008bv, Knapen:2016hky, CMS:2024nca, CMS:2026jia, ATLAS:2026fso}.
The CMS and ATLAS Collaborations have also searched for SVJs produced via resonant~\cite{CMS:2021dzg,ATLAS:2025kuz} and nonresonant~\cite{ATLAS:2023swa} processes. The searches for resonant production of SVJs use models with a leptophobic \PZpr boson mediator~\cite{Cohen:2015toa} and thus cannot be directly compared to this search. The ATLAS nonresonant search~\cite{ATLAS:2023swa} probes the same scalar mediator \PPhi as this search, but their signal model otherwise has significant differences, which are discussed further in Section~\ref{sec:DataSim}.

The present analysis utilizes data from proton-proton ($\Pp\Pp$) collisions recorded with the CMS detector at the CERN LHC during the 2016--2018 data-taking period.
This data set corresponds to an integrated luminosity of 138\fbinv~\cite{CMS:2021xjt,CMS:2026mqq}.
The CMS detector is described in Section~\ref{sec:cmsdet}.

The search follows a dual strategy using novel taggers to distinguish SVJs from SM jets, yielding two sets of results.
First, a supervised graph neural network (GNN) with the \PNET (PN) architecture~\cite{Qu:2019gqs} (Section~\ref{sec:SVJPNN}) is used to identify SVJs by training on particle-level information in simulated samples of signal models and background processes.
Second, the newly developed unsupervised Wasserstein normalized autoencoder (WNAE)~\cite{CMS:2025lmn} (Section~\ref{sec:SVJWNAE}) is implemented to identify SVJs as anomalous jets by learning the SM expectation for jet substructure variables.
While not as sensitive to the dark-shower dynamics of the signal models used to train the supervised jet tagger, the autoencoder is more model-agnostic and therefore has higher acceptance for SVJs from signal hypotheses not used in the supervised training.

The dominant SM background in this search arises from events where a mismeasured jet induces artificial \ptmiss.
This background predominantly consists of QCD multijet events.
Additional background contributions originate from \ttjets, \wlnjets, \znnjets, and single top quark processes, where genuine \ptmiss arises from neutrino(s) in the final state.
The SM background is estimated from the observed data using a recently developed automated machine-learning technique~\cite{CMS:2025cvw}.

The detailed background estimates are given in Section~\ref{sec:BkgdEstimation}, followed by the description of various systematic uncertainties in Section~\ref{sec:Syst}.
The results and interpretation are discussed in Section~\ref{sec:Int}, followed by a summary of the paper in Section~\ref{sec:concl}.
Tabulated results can be found in the HEPData record for this analysis~\cite{hepdata}.

\section{The CMS detector}\label{sec:cmsdet}

The CMS apparatus~\cite{CMS:2008xjf,CMS:2023gfb} is a multipurpose, nearly hermetic detector, designed to trigger on~\cite{CMS:2020cmk,CMS:2016ngn,CMS:2024aqx} and identify electrons, muons, photons, and (charged and neutral) hadrons~\cite{CMS:2020uim,CMS:2018rym,CMS:2014pgm}. Its central feature is a superconducting solenoid of 6\unit{m} internal diameter, providing a magnetic field of 3.8\unit{T}. Within the solenoid volume are a silicon pixel and strip tracker, a lead tungstate crystal electromagnetic calorimeter (ECAL), and a brass and scintillator hadron calorimeter (HCAL), each composed of a barrel and two endcap sections. Forward calorimeters extend the pseudorapidity ($\eta$) coverage provided by the barrel and endcap detectors. Muons are reconstructed using gas-ionization detectors interleaved with the layers of the steel flux-return yoke outside the solenoid. 
Events of interest are selected using a two-tiered trigger system. The first level, composed of custom hardware processors, uses information from the calorimeters and muon detectors to select events at a rate of around 100\unit{kHz} within a fixed latency of 4\mus~\cite{CMS:2020cmk}. The second level, known as the high-level trigger, consists of a farm of processors running a version of the full event reconstruction software optimized for fast processing, and reduces the event rate to a few kHz before data storage~\cite{CMS:2016ngn,CMS:2024aqx}.
More detailed descriptions of the CMS detector, together with a definition of the coordinate system used and the relevant kinematic variables, can be found in Refs.~\cite{CMS:2008xjf,CMS:2023gfb}.

\section{Signal model and simulation}\label{sec:DataSim}

Simulated samples are used to model signal and SM background processes.
To account for changes in the detector conditions, including detector upgrades and aging, separate samples are produced for each data-taking year: 2016, 2017, and 2018.

The signal model is based on Ref.~\cite{Cohen:2017pzm} and contains a dark sector with the following constituents: a bifundamental scalar boson \PPhi; two flavors ($\Nfdark = 2$) of fermionic dark quarks \PqdarkO, \PqdarkT; stable and unstable pseudoscalar dark hadrons \PpidarkDM and \Ppidark; stable and unstable vector dark hadrons \PrhodarkDM and \Prhodark; and a dark gluon \Pgdark that carries an unbroken $SU(\Ncdark)$ dark-QCD force with the number of dark colors $\Ncdark = 2$.

The effective parameter \rinv is defined as the average fraction of dark hadrons that are stable:
\begin{equation}
    \rinv = \left\langle \frac{\Nstable}{\Nstable+\Nunstable} \right\rangle
    \label{eq:rinv}
\end{equation}
and can take any value between 0 and 1, as described in Ref.~\cite{Cohen:2015toa}.

There are four free parameters in the model: the mass of the mediator \mPhi, the Yukawa coupling of the mediator to the SM and dark quarks \yukawa, the effective parameter \rinv, and the dark-hadron mass scale \mDark.
The signal model assumes that all dark hadrons have the same mass \mDark, and the dark-quark mass is set as $\mqdark = \frac{1}{2}\mDark$.
The inclusive production cross section \sigmaPhi---summing $t$-channel, associated, and pair production---is determined by \mPhi and \yukawa.
The flavor structure is taken to be diagonal with $\yijab = \yukawa \delta_{ij} \delta{ab}$, following Ref.~\cite{Cohen:2017pzm}.
The energy scale \lamDark of the dark-QCD force is specified in terms of \mDark, using the relationship from Ref.~\cite{CMS:2021dzg} that approximately maximizes the dark-hadron multiplicity: 
\begin{equation}
\lamDark = 3.2\GeV (\mDark/{\GeVns})^{0.8}. \label{eq:lambda_mass}
\end{equation}

When the unstable dark hadrons are produced, they decay promptly into SM quarks through the virtual \PPhi boson exchange.
The unstable vector dark hadrons \Prhodark decay to a pair of SM quarks of any flavor with equal branching fractions, provided that $\mDark > 2\mq$, where \mq is the SM quark mass.
The decays of the unstable pseudoscalar dark hadrons \Ppidark are described via a mass insertion~\cite{Rosiek:2017cwn, Dedes:2015twa, Marcano:2019rmk}, as the bifundamental scalar only couples to the right-handed quarks.
Consequently, unlike \Prhodark, the unstable pseudoscalar dark hadrons decay preferentially to more massive SM quarks.

The simulated signal samples are generated using \MADGRAPHFULL~\cite{Frederix:2018nkq} at leading order (LO).
The model, obtained from Ref.~\cite{Cohen:2017pzm}, is defined using the \FEYNRULES package~\cite{Alloul:2013bka} and is based on the \textsc{DMsimp} class of models~\cite{Afik:2018rxl}.
The signal model parameters \mPhi, \mqdark, and \yukawa are specified during the generation.

The showering and hadronization of \Pqdark are simulated in \PYTHIAFULL~\cite{Sjostrand:2014zea} using the HV module.
The probability of producing vector dark hadrons is set to $P_{\text{vector}} = 0.75$, following Ref.~\cite{Cohen:2015toa}.
The stable dark hadrons are simulated by varying the branching fraction of the dark hadrons to SM quarks according to the value of \rinv.
To enforce an overall $Z_2$ symmetry, events with an odd number of stable dark hadrons are rejected.

Since a simultaneous scan of the four parameters in the signal model is not practical, three two-dimensional scans are used, where \mPhi and one other parameter are varied, keeping the remaining two parameters at their benchmark values.
The benchmark values for the signal parameters are as follows: $\mPhi = 2000\GeV$, $\mDark = 20\GeV$, $\rinv = 0.3$, and $\yukawa = 1.0$.
These values are chosen based on considerations from Refs.~\cite{Cohen:2015toa} and~\cite{Cohen:2017pzm}.
We scan \mPhi from 500 to 4000\GeV with 12 values, \yukawa from 0.1 to 3.5 with 8 values, \rinv from 0 to 1 with 12 values, and \mDark with values of 1, 5, 10\GeV and then increasing to \mDarkMax in steps of 10\GeV.
This variable upper limit on \mDark arises to ensure there is sufficient multiplicity of dark hadrons to constitute a shower.
To achieve this, the confinement scale should be much smaller than the mediator mass, typically~\cite{Albouy:2022cin}:
\begin{equation}
\mPhi \gtrsim 30 \lamDark.
\label{eq:shower_cond}
\end{equation}
By combining Eq.~\eqref{eq:shower_cond} with Eq.~\eqref{eq:lambda_mass}, the \mDarkMax value can be found for a given \mPhi:
\begin{equation}
\mDarkMax = \left(\frac{\mPhi}{96\GeV}\right)^{1.25}\GeV.
\label{eq:mdarkmax}
\end{equation}

The \znnjets, \wlnjets, \ttjets, and single top quark backgrounds are generated with \MGvATNLO~\cite{Alwall:2014hca,Alwall:2007fs} at LO and hadronized with \PYTHIA~\cite{Sjostrand:2014zea}.
The event generation includes up to four additional partons in the matrix element calculations for \ttjets and three additional partons for \zjets and \wjets.
These samples are normalized to the next-to-next-to-LO (NNLO) cross sections~\cite{Beneke:2011mq,Cacciari:2011hy,Baernreuther:2012ws,Czakon:2012zr,Czakon:2012pz,Czakon:2013goa,Li:2012wna}, and the MLM scheme~\cite{Alwall:2007fs} is used to match jets from the matrix element calculations and the parton shower.
The QCD multijet background samples are generated at LO and hadronized with \PYTHIA.

For all signal and background samples, the NNPDF3.1 NNLO parton distribution functions (PDFs)~\cite{Ball:2017nwa} are used, with the underlying event tune CP5~\cite{Sirunyan:2019dfx}.
All events pass through the full CMS detector simulation using \GEANTfour~\cite{Agostinelli:2002hh}, which reflects the state of the detector for each data-taking year, including any upgrades.
Multiple simultaneous $\Pp\Pp$ interactions (pileup) are also included in the simulation.
The simulated samples are corrected to make the pileup distribution match the distribution in data as closely as possible.

\section{Event reconstruction and selection}\label{sec:EvtReco}

A particle-flow (PF) algorithm~\cite{CMS:2017yfk} aims to reconstruct and identify each individual particle in an event, with an optimized combination of information from the various elements of the CMS detector. The energy of photons is obtained from the ECAL measurement. The energy of an electron is determined from a combination of the electron momentum at the primary interaction vertex as determined by the tracker, the energy of the corresponding ECAL cluster, and the energy sum of all bremsstrahlung photons spatially compatible with originating from the electron track. The energy of a muon is obtained from the curvature of the corresponding track. The energy of a charged hadron is determined from a combination of the momentum measured in the tracker and the matching ECAL and HCAL energy deposits, corrected for the response function of the calorimeters to hadronic showers. Finally, the energy of neutral hadrons is obtained from the corresponding corrected ECAL and HCAL energies.

For each event, hadronic jets are clustered from these reconstructed particles (PF candidates) using the infrared- and collinear-safe anti-\kt algorithm~\cite{Cacciari:2008gp, Cacciari:2011ma}.
Jet momentum is determined as the vectorial sum of all particle momenta in the jet, and is found from simulation to be, on average, within 5 to 10\% of the true momentum over the entire transverse momentum (\pt) spectrum and detector acceptance.
The \ptmiss vector \ptvecmiss is computed as the negative vector \pt sum of all the PF candidates in an event~\cite{CMS:2019ctu}. The \ptvecmiss is modified to account for corrections to the energy scale of the reconstructed jets in the event.
To enhance the robustness of the reconstruction, electrons~\cite{CMS:2020uim} and muons~\cite{CMS:2018rym} are required to have $\pt > 10\GeV$ and $\abseta < 2.4$ and to satisfy additional quality criteria.
The primary vertex (PV) is taken to be the vertex corresponding to the hardest scattering in the event, evaluated using tracking information alone, as described in Section 9.4.1 of Ref.~\cite{CMS-TDR-15-02}.

This analysis utilizes two distinct categories of jets.
The primary jet collection consists of jets with $\pt>50\GeV$ and $\abseta < 2.4$, clustered with a distance parameter of $R = 0.8$.
This choice is motivated by the expectation that signal jets exhibit a wider radiation pattern compared to those originating from SM background processes.
These reconstructed objects are denoted by a capital \widejet and are referred to simply as ``jets'' throughout this paper.
The pileup-per-particle identification algorithm~\cite{CMS:2020ebo,Bertolini:2014bba} is used to mitigate the effect of pileup on these jets at the reconstructed particle level, making use of local shape information, event pileup properties, and tracking information. A local shape variable is defined, which distinguishes between collinear and soft diffuse distributions of other particles surrounding the particle under consideration. The former is attributed to particles originating from the hard scatter and the latter to particles originating from pileup interactions. Charged particles identified to be originating from pileup vertices are discarded. For each neutral particle, a local shape variable is computed using the surrounding charged particles compatible with the PV within the tracker acceptance ($\abseta < 2.5$), and using both charged and neutral particles in the region outside of the tracker coverage. The momenta of the neutral particles are then rescaled according to their probability to originate from the PV, deduced from the local shape variable, superseding the need for jet-based pileup corrections~\cite{CMS:2020ebo}.
Jet energy corrections are derived from simulation studies so that the average measured energy of jets becomes identical to that of particle level jets. In situ measurements of the momentum balance in dijet, $\PGg+\text{jet}$, $\PZ+\text{jet}$, and multijet events are used to determine any residual differences between the jet energy scale in data and in simulation, and appropriate corrections are made~\cite{CMS:2016lmd}. In simulated events, jet \pt values are smeared according to measured resolutions to ensure consistency with the observed data. Additional selection criteria are applied to each jet to remove jets potentially dominated by instrumental effects or reconstruction failures~\cite{CMS:2020ebo}.

A secondary collection of jets is clustered with a distance parameter of $R=0.4$.
These jets are required to satisfy $\pt>30\GeV$ and $\abseta < 2.4$; they are denoted by a lowercase \narrowjet and referred to as ``narrow jets'' throughout the paper.
Narrow jets are used for computing \HT, defined as the scalar \pt sum of all narrow jets, and to mitigate instrumental backgrounds.
To mitigate the effect of pileup on these jets, tracks identified to be originating from pileup vertices are discarded and an offset correction is applied to correct for remaining contributions~\cite{CMS:2020ebo}.
The \ptvecmiss is modified to account for corrections to the energy scale of the narrow jets in the event.
Jets originating from bottom quarks are identified via the \textsc{DeepJet} algorithm~\cite{Bols:2020bkb}.
Narrow jets passing the tight \PQb tagging working point, defined to accept 0.1\% of light-flavor quark and gluon background jets, are denoted as \PQb jets and are used to identify control regions (CRs) to train the unsupervised SVJ tagger, described in Section~\ref{sec:SVJWNAE}.

Several data sets are utilized in this paper, collected using triggers that place requirements on the jet \pt, the \ptmiss, the \HT, or the \mht, which is the magnitude of the negative vector sum of the narrow-jet \pt.
These data sets are combined to keep each selected event only once, regardless of how many triggers it passed.
The most efficient triggers require jet $\pt > 450\,(500)\GeV$ or $\HT > 800\,(1050)\GeV$ for data taken during 2016 (2017--2018).
The next most efficient trigger selects events with $\ptmiss > 110\,(120)\GeV$ and $\mht > 120\,(120)\GeV$ for 2016 (2017--2018).
The final trigger requires events to have $\HT > 300\,(500)\GeV$, $\ptmiss>110\,(100)\GeV$, and $\mht > 0\,(75)\GeV$ for 2016 (2017--2018).

Electrons and muons are used for the lepton veto definition and corresponding CR identification, called ``veto leptons'', and also to identify the top quark jet in a top-quark-enriched CR used for the unsupervised SVJ tagger training, called ``tag leptons''.
Veto and tag leptons are required to have $\pt > 10\GeV$ and to have $\abseta < 2.4$ in order to fall within the tracker acceptance region.
They are also required to satisfy, respectively, loose and medium quality selection criteria as defined in Refs.~\cite{CMS:2020uim,CMS:2023dsu}.
To reduce the contributions from jets misidentified as leptons or leptons originating from hadron decays rather than from prompt {\PW} boson decays, an isolation requirement is applied to both veto and tag leptons.
The ``mini-isolation''~\cite{CMS:2023dsu} is defined as $\Imini(\ell) = (\sum_{\dRmax}{\ptsub{\Pchhad}}+\text{max}[0,\sum_{\dRmax}{\ptsub{\PGg}+\ptsub{\Pnhad}-\ptsub{\text{pileup}}/2}])/\ptsub{\ell}$,
where the different particle candidates are denoted by $\ell$ for the charged lepton, \Pchhad for charged hadrons, \PGg for photons, and \Pnhad for neutral hadrons.
The label ``pileup'' in the last term denotes charged hadrons that do not originate from the PV and are used to estimate additional energy coming from neutral particles whose vertex cannot be determined.
All the sums consider candidates within a cone of $\dR = \sqrt{\smash[b]{(\Deta)^2+(\Dphi)^2}} < \dRmax$,
where $\phi$ is the azimuthal angle in radians, and \dRmax is 0.2 for $\ptsub{\ell} < 50\GeV$, $10\GeV/\ptsub{\ell}$ for $50 < \ptsub{\ell} < 200\GeV$, and 0.05 for $\ptsub{\ell} > 200\GeV$.
The radius of the cone decreases with increasing charged-lepton \pt in order to account for the effect of the Lorentz boost of the lepton's parent particle; a larger boost leads to more collimated decay products~\cite{Rehermann:2010vq}.
Tag electrons and muons, and veto electrons are required to satisfy $\Imini(\ell) < 0.1$, whereas veto muons must satisfy $\Imini(\Pgm) < 0.4$.

\begin{figure}[htb!]
\centering
\includegraphics[width=0.49\linewidth]{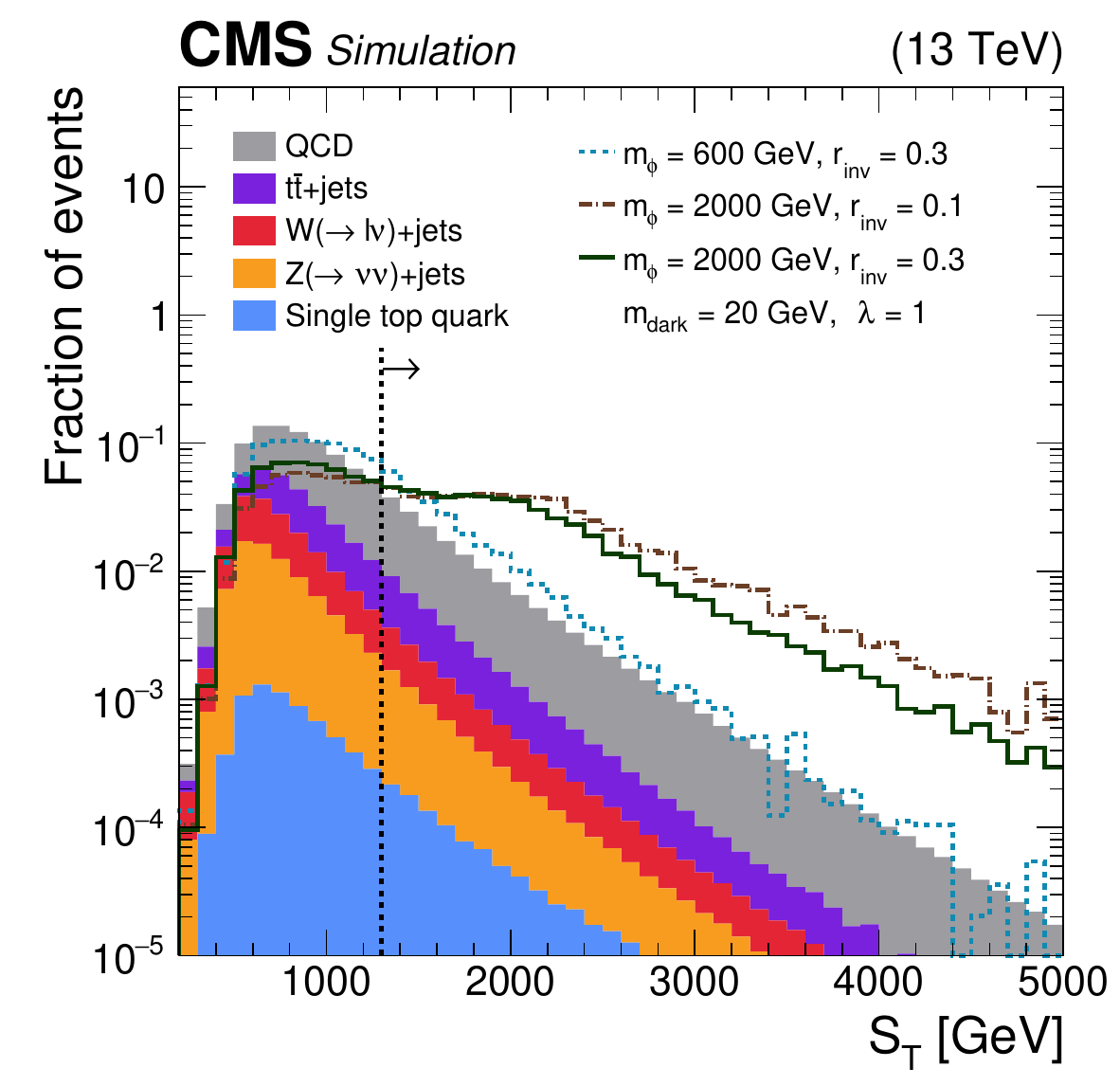}
\includegraphics[width=0.49\linewidth]{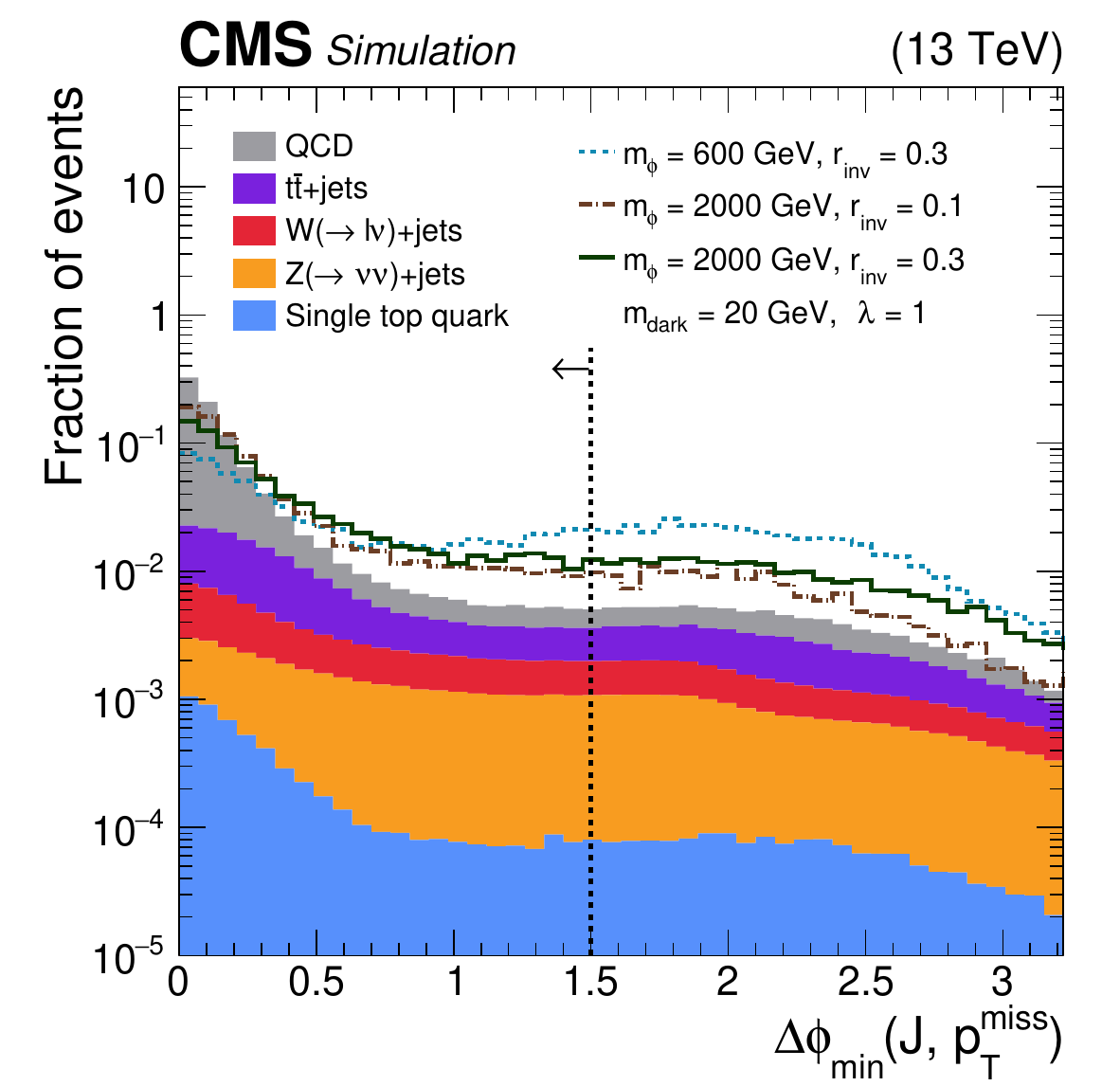}
\includegraphics[width=0.49\linewidth]{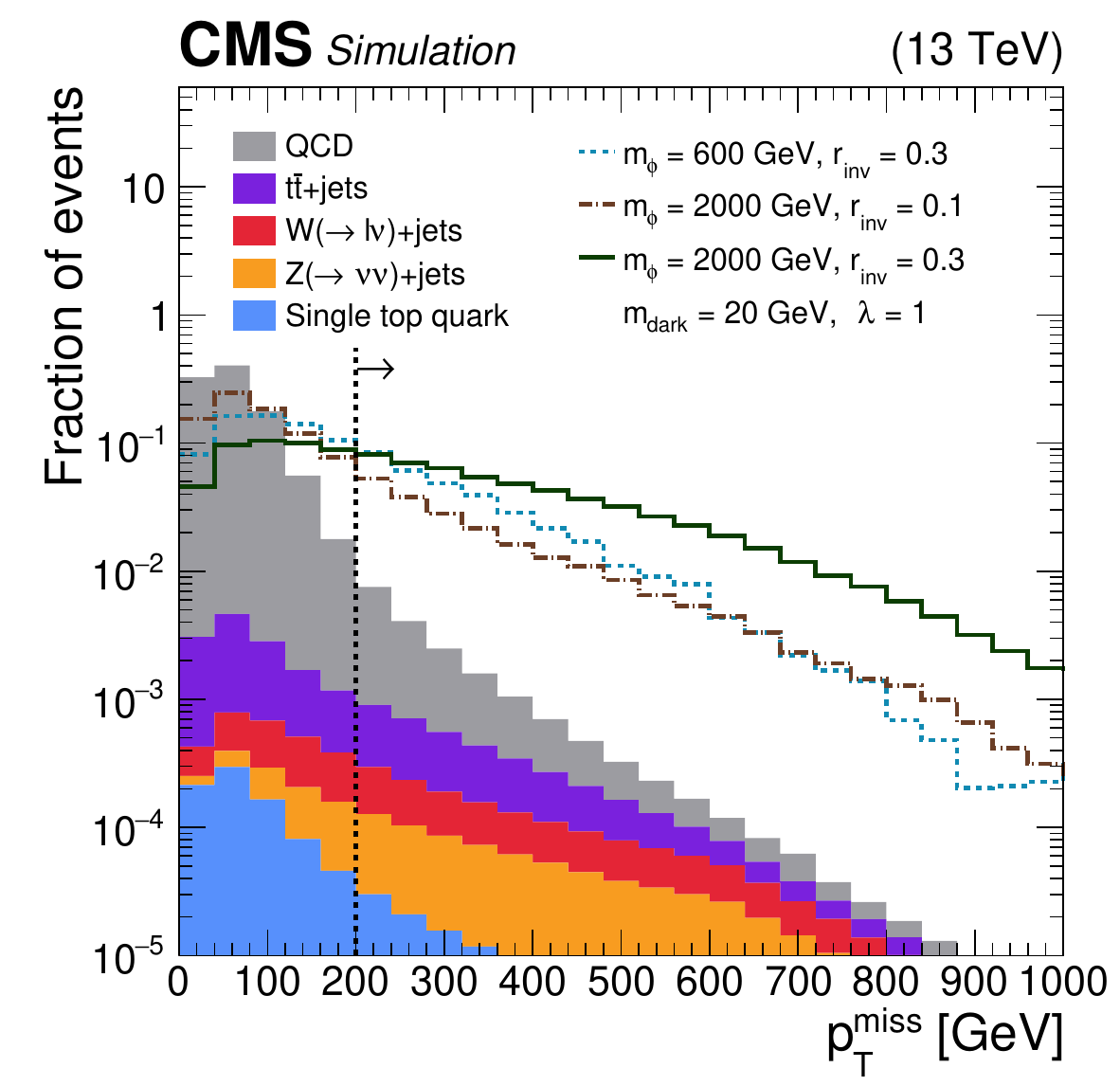}
\caption{Unit-normalized distributions of the discriminating variables \ST, \mindphi, and \ptmiss for the simulated SM backgrounds and several signal models. For each variable, all preselection criteria are applied except the selection on that variable. The vertical black dotted line indicates the preselection requirement on the variable, with the arrow indicating the region of the distribution included in the analysis.}
\label{fig:preselection}
\end{figure}

The preselection criteria for this analysis are optimized to enhance signal sensitivity, while suppressing background contributions.
The scalar sum of transverse momenta $\ST = \HT + \ptmiss$ is required to be greater than 1300\GeV to select events in the trigger efficiency plateau across all three years of data taking.
Events are required to contain at least two jets \widejet, reflecting the expected signal topology characterized by the presence of multiple high-energy jets in the final state.
To further suppress background contributions, events with identified and mini-isolated electrons or muons are vetoed.
This criterion significantly reduces background contamination from \wjets and \ttjets processes, where \ptmiss primarily originates from neutrinos produced in leptonic \PW boson decays.
To suppress backgrounds from electroweak processes, the minimal angle between all jets and the \ptvecmiss, \mindphifull, is required to be less than 1.5, as SVJ events are expected to have \ptmiss aligned with at least one of the jets in the event.
Additionally, a minimum \ptmiss requirement of 200\GeV is imposed to eliminate low-\ptmiss QCD events.
The distributions of \ST, \mindphi, and \ptmiss are shown in Fig.~\ref{fig:preselection} for background and a selection of signal hypotheses.

Additional criteria are applied to reject events with anomalously high \ptmiss values from reconstruction failures, detector malfunctions, or other noncollision backgrounds.
A large fraction of these events are rejected by dedicated filters designed to identify at least 85\% of the anomalous events, while having a false-positive rate less than 0.1\%~\cite{CMS:2019ctu}.
The leading source of instrumental background is mismeasurement of jet energy due to a small number of nonfunctional ECAL readout channels.
A dedicated filter, optimized for the kinematic phase space of this search, is employed to identify the nonfunctional channels, following the approach in Ref.~\cite{CMS:2021dzg}.
A separate population of events with anomalously high \ptmiss arises from rare failures in specific ECAL reconstruction algorithms that lead to misreconstructed jets.
These events were identified at the preselection level, and are rejected by vetoing events where the leading narrow jet has a photon energy fraction $f_{\PGg} > 0.7$ and $\pt > 1.0\TeV$, as in Ref.~\cite{CMS:2021dzg}.
In the later portion of the 2018 data-taking period, corresponding to 38.65\fbinv, a section of the HCAL was not active.
This effect is not included in the detector simulation, so events in which any narrow jet falls into the inactive section ($-3.05 < \eta < -1.35$, $-1.62 < \phi < -0.82$) are rejected.

The signal simulation is corrected to account for mismodeling of the SVJs, as the analysis is sensitive to changes in jet substructure.
This correction uses a modified version of the method in Ref.~\cite{CMS:2025eyd}, with the corresponding statistical and systematic uncertainties, discussed in Section~\ref{sec:Syst}.
The method reclusters the constituents of simulated signal jets into subjets, each corresponding to a single SM quark, and corrects the modeling of these subjets based on their Lund jet plane density~\cite{Dreyer:2018nbf} by comparing to a data CR.
This allows multi-pronged exotic objects to be corrected based on SM data.
The data CR selects hadronic \ttbar events, where the single quark subjets are then taken from the hadronic \PW boson decay.
In this analysis, to obtain quark subjets from the signal jets, the generator-level information in the signal samples is used to match each dark hadron, an SM color singlet, to its final-state decay products.
These generator-level final-state particles are then matched to the reconstruction-level PF candidates that comprise the jet constituents.
The set of jet constituents corresponding to each dark hadron is exclusively clustered into two subjets, because the dark hadrons decay to \qqbar pairs.
The correction is then derived, as described in Ref.~\cite{CMS:2025eyd}, using the ratio of the Lund jet plane density between the simulated and observed subjets of the jets from hadronically decaying \PW bosons.
This correction is used to reweight the simulated signal events, resulting in a change in the yield of up to 8.2\%, where signal models with extremely low \yukawa values see a larger change in the yield compared to models with $\yukawa = 1$.

After the preselection is applied, the QCD multijet production contributes 65\% of expected events, while the \ttjets process contributes 22\%, \wjets 7.1\%, \zjets 5.6\%, and single top quark production 0.96\%.
Contributions from diboson production are found to be negligible.
The signal hypotheses with higher mediator masses show a slightly increased \ST relative to the aggregated background.
The signal \ptmiss is significantly larger compared to that for the background, consistent with the nonresonant nature of the signal process, providing the best discrimination power.
Therefore, this analysis searches for an excess of events in the \ptmiss distribution.

\section{Identification of SVJs}\label{sec:SVJtaggers}

Two strategies are utilized to identify SVJs: supervised and unsupervised machine-learning taggers.
The supervised tagger learns from the simulated signals how to characterize the SVJs and differentiate them from SM jets.
This method is dependent on specific signal model details, primarily the DS parameters and the dark showering, but shows the achievable sensitivity from fully exploiting the characteristics of the SVJs.
In contrast, the unsupervised tagger is agnostic to the details of the dark shower, as the tagger is trained on data corresponding to SM background processes and thereby learns to identify any jets with anomalous substructure.
This tagger forgoes some discriminating power for any particular signal model, but remains effective across a much broader range of possible signal models.

For both methods, jets are tagged as SVJs if the output score of the tagger is above a given threshold.
Events are then categorized based on the number of tagged SVJs (\nsvjpn or \nsvjwnae for the \strategies, respectively) into four categories: $\nsvj^{i} = 0$, 1, 2, and $\geq3$.
Both methods are designed to reject the backgrounds that form the largest fraction of events after the preselection described in Section~\ref{sec:EvtReco}, specifically in the signal region (SR) described in Section~\ref{sec:BkgdEstimation}.
These backgrounds are the production of jets from QCD interactions at 51\% and \ttjets with 20\%, which together provide 71\% of the total background.
Additional background contributions in the SR come from \wjets (15\%), \zjets (11\%), and single top quark production (3\%).

\subsection{Supervised tagger}\label{sec:SVJPNN}

Traditional jet substructure observables, developed to identify jets from weak decays of Lorentz-boosted heavy resonances, are not optimal for distinguishing between SM jets and SVJs.
Therefore, an advanced neural network architecture, \PNET~\cite{Qu:2019gqs} is employed to develop an SVJ tagger that uses fundamental information such as the jet constituent four-vectors as inputs.

In this case, a jet is represented by a particle cloud: an unordered, permutationally invariant set of points, each carrying a vector that contains the constituent particle features.
The \PNET algorithm has a GNN architecture that can operate on this point cloud data structure.
The first 100 constituents of each jet, ordered by \pt, are used as inputs.
The input variables are, for each constituent: $\Deta$, $\Dphi$, and $\dR$ with respect to the jet axis; $\log_{10}{(\pt/{\GeVns})}$, $\log_{10}{(E/{\GeVns}})$, $\log_{10}{(\pt/\pt(\widejet))}$, $\log_{10}{(E/E(\widejet))}$, and the reconstructed particle type.

\begin{figure}[ht!]
	\centering
	\includegraphics[width=0.4\linewidth]{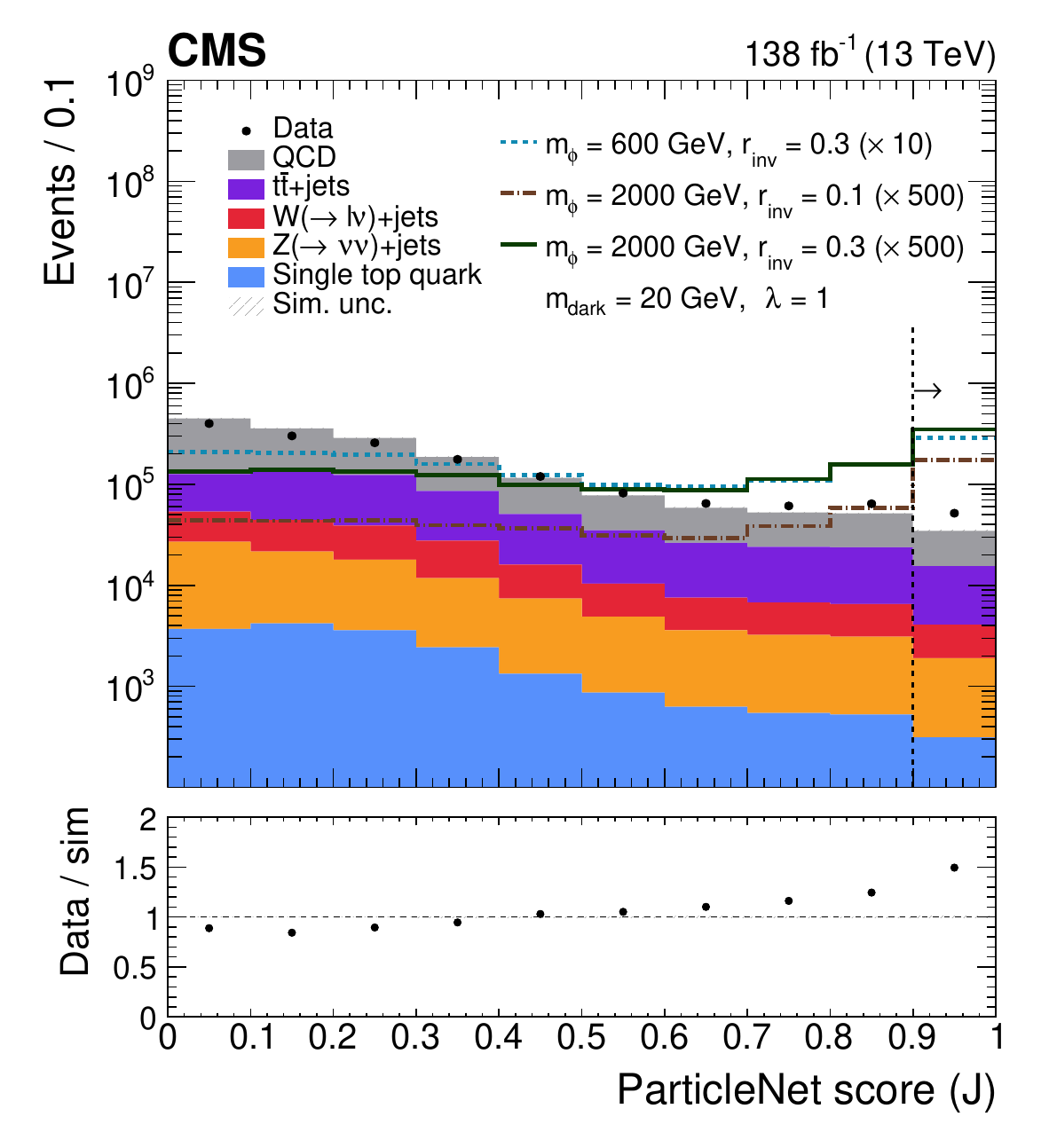}
    \includegraphics[width=0.58\linewidth]{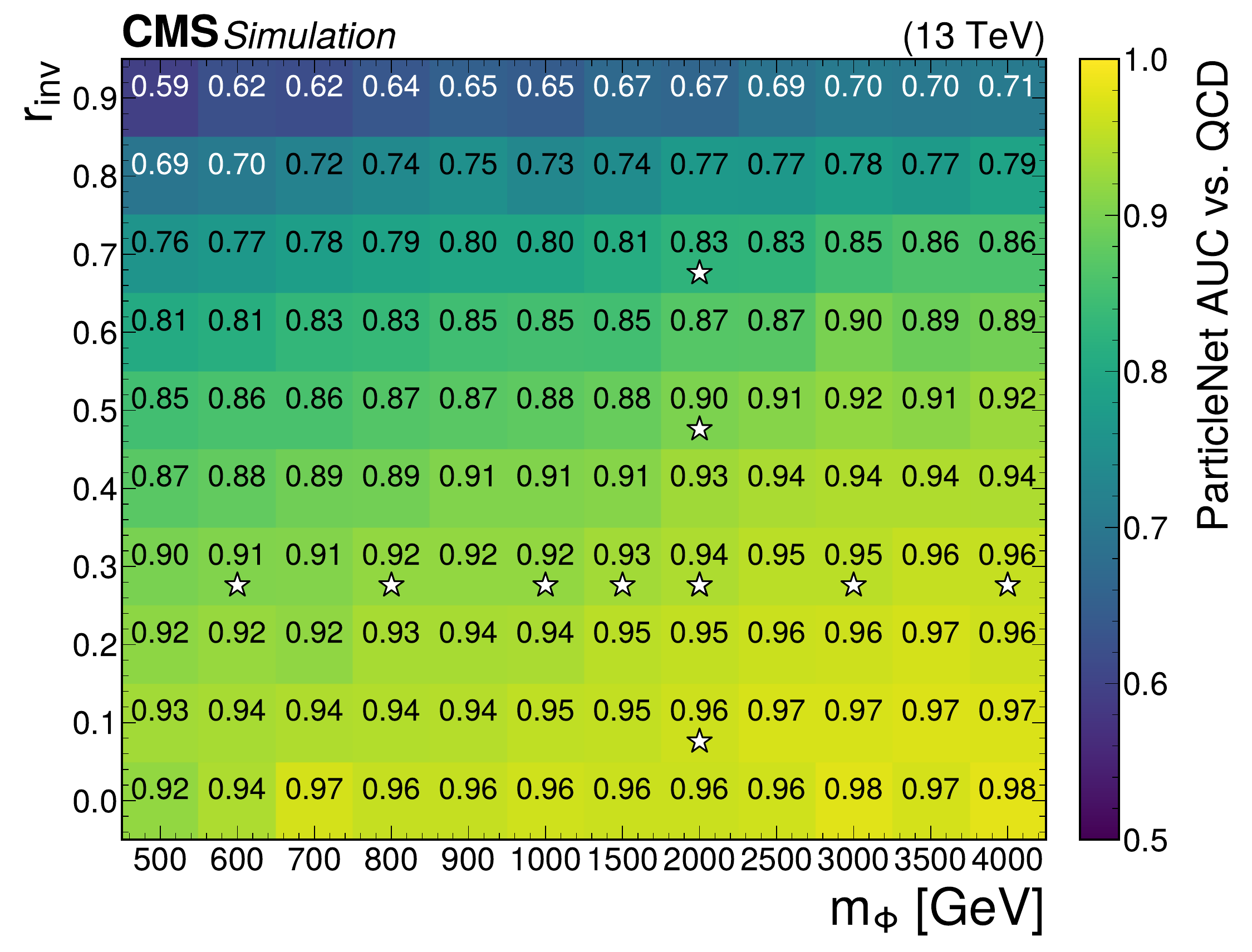}
	\caption{The \PNET score distribution for data, background simulation, and a selection of signal models scaled for visibility (left), and the performance discriminating between the simulated QCD background and signal models with varying \mPhi and \rinv values with $\mDark = 20\GeV$ and $\yukawa = 1$ (right), after applying the preselection criteria. The lower panel shows the ratio between the data and the simulated background. The signal models used for training \PNET are marked with a star. The simulated background score distributions are shown for illustration purposes only; they are not used to predict the SM background in the statistical interpretation. The signal jets are corrected for substructure-related mismodeling using their Lund jet plane density, as described in Section~\ref{sec:EvtReco}. The background simulation has no substructure correction applied. The statistical uncertainties in the data points are smaller than the marker size. The statistical uncertainty in the background simulation from the limited number of simulated events is shown, but is generally small.}
	\label{fig:taggerPNET}
\end{figure}

To train the network, 12 signal points are used, shown with the stars in Fig.~\ref{fig:taggerPNET} (right): 7 variations of \mPhi (600, 800, 1000, 1500, 2000, 3000, 4000\GeV), 4 of \mDark (1, 20, 50, 100\GeV), and 4 of \rinv (0.1, 0.3, 0.5, 0.7).
Each parameter is varied separately, starting from the benchmark signal model.
Each signal model contributes between $\approx$10\,000 and $\approx$30\,000 jets to the training data set.
A subset of the QCD multijet and \ttjets samples are used as backgrounds, such that the training is performed with an equal number (${\approx}3\times10^5$) of light-flavor QCD, top quark, and signal jets.
The full data set of background and signal samples as described above is split into training (70\%), testing (15\%), and validation (15\%) sets.
The samples corresponding to the 2018 data-taking conditions are used for training, using jets from the events that pass the preselection in Section~\ref{sec:EvtReco}.
The tagger has similar performance across the different data-taking years.

The neural network score can have a significant dependence on the jet \pt, where jets with lower \pt tend to have higher neural network scores.
To prevent the neural network from using the jet \pt as a discriminant, jets are randomly selected in each background sample to match the \pt distribution of the signal jets, such that the final \pt spectrum in the training data is identical between backgrounds and signal.
This is done for the training, validation, and testing samples, still keeping equal numbers of jets from QCD multijet production, \ttjets, and signal.

In addition, the signal jets are required to have been produced by dark quarks or dark gluons, either from the mediator or from radiation.
At the generator level, each signal jet must have at least one constituent that can be traced back to a dark-hadron parent.
Other jets in signal events that contain SM quarks from the mediator are rejected, even if they overlap with a jet that has a constituent with a dark-hadron parent.
This selection ensures the training is performed on signal jets that have kinematic behavior distinct from SM jets.

The distribution of the \PNET score for data, background simulation, and a selection of signal models is shown in Fig.~\ref{fig:taggerPNET} (left).
The discrimination power of the tagger is quantified via the area under the curve (AUC) of the receiver operating characteristic (ROC) computed from the score.
The \PNET tagger AUC values for a selection of signal models with $\mDark = 20\GeV$ and $\yukawa = 1$ for varying \mPhi and \rinv are shown in Fig.~\ref{fig:taggerPNET} (right).
While there is some variation depending on the signal model parameters, overall the tagger performs well, especially for signal hypotheses with lower values of \rinv and higher mediator masses. 
The tagger performs best, with AUCs around 0.9, for \mDark values of 20--50\GeV, while still performing well, with AUCs near 0.8, for other values.
For signal models with the extreme low value $\mDark=1\GeV$, the AUCs are around 0.65.

Jets are tagged as SVJs if the \PNET score is above a certain threshold.
This threshold was chosen by considering a joint optimization of the \PNET threshold choice, the ABCD boundaries, and the event classifier threshold, both described in Section~\ref{sec:BkgdEstimation}, while ensuring a reasonable background efficiency.
A very high threshold on the \PNET score (0.9) was chosen to maximize performance, resulting in very few background events passing this selection in the $\nsvjpn \geq 3$ category.
This \PNET score threshold gives a jet-level signal efficiency of 34\% for the baseline signal hypothesis ($\mPhi = 2\TeV$, $\mDark = 20\GeV$, $\yukawa = 1$) and a jet-level background efficiency of 2\%.

\subsection{Unsupervised tagger}\label{sec:SVJWNAE}

The parton showering and hadronization dynamics in dark-QCD models depend on a large number of parameters, such that even if dark QCD is realized in nature, any specific signal hypothesis is unlikely to choose correct values for all of the model parameters.
Therefore, an \unsupervised for SVJ tagging is outlined in this section, complementary to the one presented in Section~\ref{sec:SVJPNN}.
The \unsupervised does not depend on direct knowledge of the jet substructure of the signal model; instead, when trained on data, it learns the probability distribution of the SM background to act as a model-agnostic anomaly detector.
The WNAE architecture~\cite{CMS:2025lmn} was developed specifically for this search by modifying the traditional normalized autoencoder~\cite{pmlr-v139-yoon21c} training objective to improve robustness and achieve better, more consistent performance.
Crucially, the WNAE is designed to map the probability distribution of the training data set directly to the reconstruction error of the autoencoder, rather than simply learning to compress the input data as a traditional autoencoder does, which improves the ability of the tagger to detect anomalous jets outside of the learned distribution.
The WNAE is a deep neural network with input and output layers of equal dimensionality.
The mean squared error between the input and output of the network is used as an anomaly score.
During training, the WNAE is tasked to learn the probability distribution of the training data set by minimizing an optimal transport criterion, assigning higher anomaly scores to events with lower probability.
If trained on a data sample devoid of signal, the WNAE can thus detect anomalies during inference, discriminating between SVJs and SM jets in a fully unsupervised manner, having not been trained on any information about the signal model.

The WNAE utilized in this search is a fully connected network of 11 hidden layers, with 32 nodes each.
It takes as input a set of 17 variables describing the internal structure of the jet: the minor and major axes~\cite{CMS:2013kfa}; the \pt dispersion \ptD~\cite{CMS:2013kfa}; the soft-drop mass~\cite{Larkoski:2014wba}; the 2-, 3-, 4-, and 5-subjettiness $\tau_{2\text{--}5}$~\cite{Thaler:2010tr}; the $C_2^{\beta=1}$, $C_2^{\beta=2}$, $D_2^{\beta=1}$, $M_2^{\beta=1}$, and $N_2^{\beta=1}$ energy correlation functions~\cite{Moult:2016cvt}; and the fractions of the total jet energy carried by charged hadrons, neutral hadrons, electrons, and photons.

Two training regions (TRs) are defined in the observed data, one enriched in QCD multijet events and another enriched in \ttjets events.
The QCD-enriched region is built by selecting \PGg+\widejet events, requiring the presence of a photon with $\pt > 200\,(250)\GeV$ for data recorded in 2016 (2017--2018), $\ST>800\GeV$, no veto lepton, and at least one jet with an angular separation from the highest \pt photon $\dR(\PGg,\widejet)>0.8$.
The \ttbar-enriched region is defined by selecting events with $\ST > 1.1\TeV$, exactly one tag lepton, no additional veto lepton, at least four narrow jets, of which exactly two are \PQb tagged, and at least one wide jet.
In the \ttbar-enriched TR, all narrow (wide) jets are required to have an angular separation from the tag lepton $\dR(\ell,\narrowjet(\widejet)) > 0.4 (0.8)$.
The expected yield for most signal hypotheses in these TRs is found to be well below 0.1\% of the SM background, resulting in a very low level of signal contamination that does not degrade the anomaly detection capabilities of the WNAE.
The data collected during 2018 are used for the training, while the other data-taking years are used to validate that this approach generalizes well.

As the jet substructure depends on the \pt of the jet, the WNAE can implicitly learn the \pt distribution of the training data set, thus tagging jets as anomalous based on their \pt rather than their internal structure.
To mitigate this effect, jets in the TRs are separated into five \pt bins $[0, 200)$, $[200, 300)$, $[300, 400)$, $[400, 500)$, and $>500\GeV$.
Five separate WNAE models are trained, one for each \pt bin, and each model is subsequently evaluated only for jets in the corresponding \pt range.
The training data set is built to contain 20\% of jets from the QCD-enriched TR and 80\% from the \ttbar-enriched TR.
This choice is motivated by the fact that, while the QCD multijet background is the most abundant, the \ttbar background is more similar to the signal.
Increasing the fraction of \ttbar jets in the training data set was found to improve the performance of the WNAE against this background, while not degrading the performance against the QCD multijet background.

\begin{figure}[ht!]
	\centering
	\includegraphics[width=0.4\linewidth]{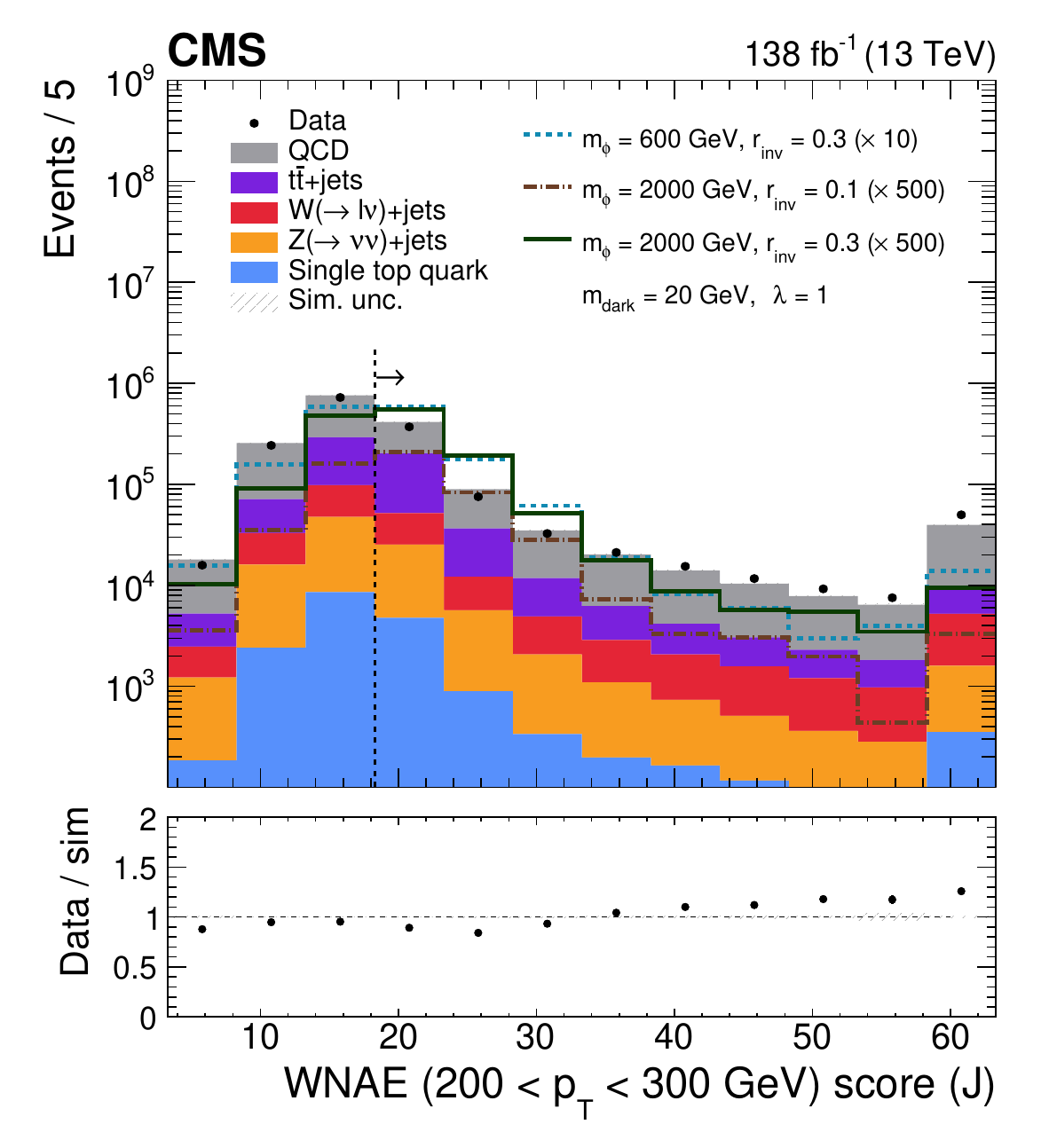}
    \includegraphics[width=0.58\linewidth]{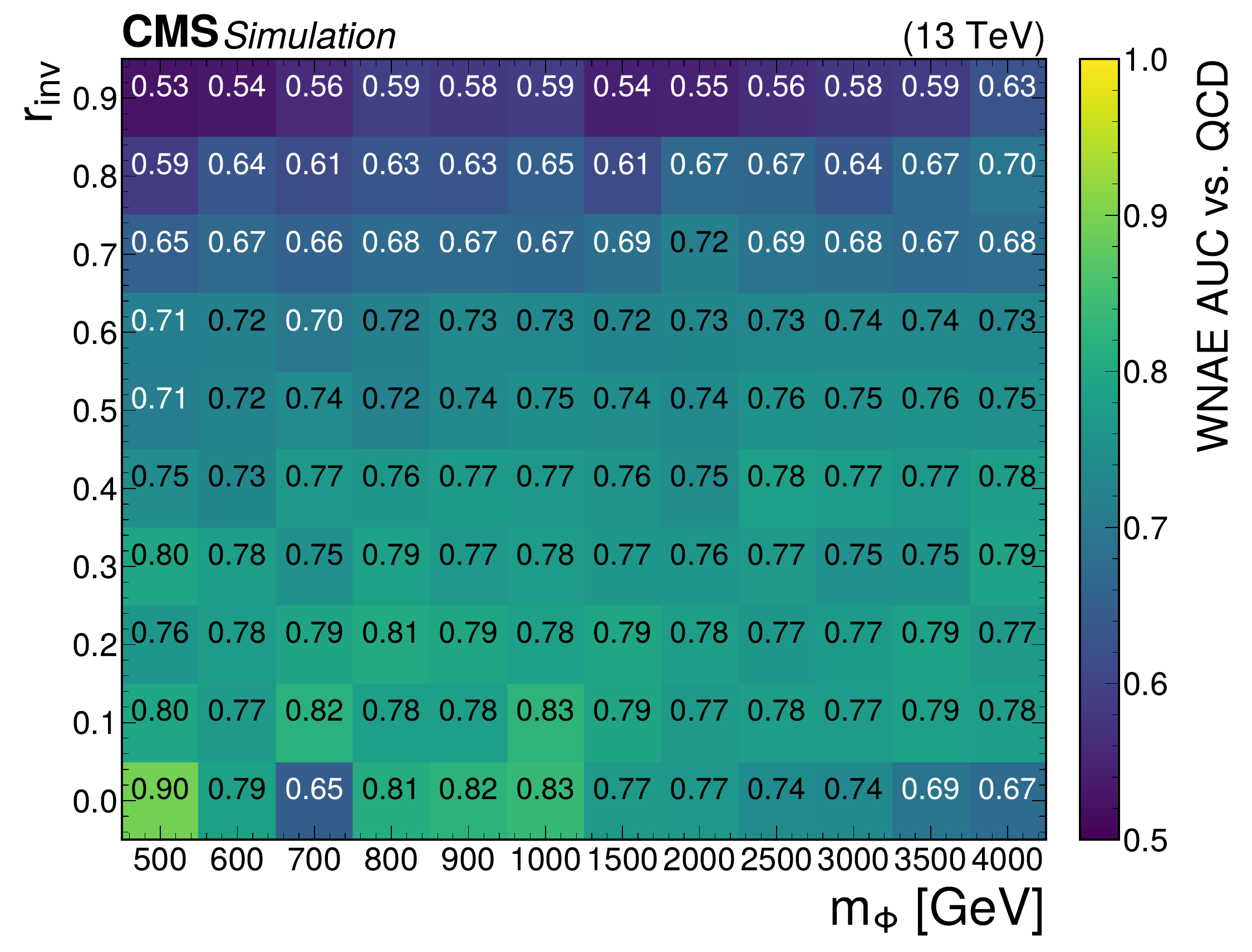}
	\caption{The WNAE score distribution for data, background simulation, and a selection of signal models scaled for visibility (left), and the performance discriminating between the simulated QCD background and signal models with varying \mPhi and \rinv values with $\mDark = 20\GeV$, $\yukawa = 1$ (right), after applying the preselection criteria, for the model trained on jets with $200 < \pt < 300\GeV$. The lower panel shows the ratio between the data and the simulated background. The fluctuations of the WNAE AUC at low \rinv and \mPhi arise from the limited number of events in this region. The simulated background score distributions are shown for illustration purposes only; they are not used to predict the SM background in the statistical interpretation. The signal jets are corrected for substructure-related mismodeling using their Lund jet plane density, as described in Section~\ref{sec:EvtReco}. The background simulation has no substructure correction applied. The statistical uncertainties in the data points are smaller than the marker size. The statistical uncertainty in the background simulation from the limited number of simulated events is shown, but is generally small.}
	\label{fig:taggerWNAE}
\end{figure}

The WNAE score distributions and the ROC AUC for signal hypotheses with varying \mPhi and \rinv values are shown in Fig.~\ref{fig:taggerWNAE}. The WNAE displays good discrimination, especially for lower values of \rinv, considering that the tagger has no access to specific signal models during training.

Jets are tagged as SVJs by the WNAE if their anomaly scores are above a given threshold. The thresholds are chosen to achieve the same background efficiency across all \pt bins in order to preserve the performance of the background estimation, and are optimized to maximize the expected significance improvement for signals in the low \rinv range, since signals at high \rinv are expected to be more easily identified based on their \ptmiss.
The thresholds are thus chosen to give a background jet-level efficiency of 20\% in each \pt bin. The corresponding signal jet-level efficiency is 67\% for the baseline signal hypothesis ($\mPhi = 2\TeV$, $\mDark = 20\GeV$, $\rinv = 0.3$, $\yukawa = 1.0$).

\section{Background estimation}\label{sec:BkgdEstimation}

Given that the signal production is dominated by nonresonant processes, this analysis searches for an excess in the \ptmiss distribution of candidate events.
The background composition at the preselection level in Section~\ref{sec:EvtReco} changes when requiring different numbers of tagged SVJs.
While QCD multijet production is still the primary background, its fraction decreases from 83\% for $\nsvjpn = 0$ to 48\% for $\nsvjpn \geq 3$, with other contributions increasing correspondingly.
For the WNAE tagger, the QCD background fraction also decreases, but less dramatically in categories with higher \nsvjwnae.
The full list of the background fractions in the different \nsvj categories for both taggers is shown in Table~\ref{tab:bkgFraction}.
The signal sensitivity generally improves as the required number of tagged SVJs increases.

\begin{table}[!hbtp]
\topcaption{The simulated background fractions [\%] after applying the preselection criteria in the different \nsvj categories for the \PNET and WNAE SVJ taggers.}
\centering
\renewcommand{\arraystretch}{1.1}
\begin{tabular}{l c c c c c c c c }
& QCD & \ttjets & \wlnjets & \znnjets & Single top quark \\
\hline
$\nsvjpn = 0$ & 82 & 6 & 6 & 5 & 1 \\
$\nsvjpn = 1$ & 78 & 10 & 6 & 5 & 1 \\
$\nsvjpn = 2$ & 75 & 15 & 5 & 4 & 1 \\
$\nsvjpn \geq 3$ & 48 & 42 & 2 & 8 & 0 \\ [1.5\cmsTabSkip]
$\nsvjwnae = 0$ & 83 & 7 & 5 & 4 & 1 \\
$\nsvjwnae = 1$ & 79 & 10 & 6 & 4 & 1 \\
$\nsvjwnae = 2$ & 75 & 14 & 7 & 3 & 1 \\
$\nsvjwnae \geq 3$ & 73 & 17 & 7 & 3 & 1 \\
\end{tabular}
\label{tab:bkgFraction}
\end{table}

As shown in Table~\ref{tab:bkgFraction}, the primary source of background in all \nsvj categories remains the QCD multijet production, which can mimic the signal signature when the energy of a jet is mismeasured, leading to artificial \ptmiss.
As an instrumental background, it is especially challenging to estimate reliably in simulation, necessitating an approach based on control samples in data.
The ABCDisCoTEC~\cite{CMS:2025cvw} method is employed to estimate the background from the observed data and to maximize the separation between signal and SM backgrounds using event kinematic variables.
This technique is based on the ABCD method~\cite{CDF:1990kbl,Kasieczka:2020pil,Buttinger:2018,Behnke:2013pga}, which utilizes two discriminating independent variables $x$ and $y$ to estimate the background in the SR from a set of three CRs.
These regions are defined by placing a threshold value in each of the two variables, such that the $x$-$y$ plane is divided into four independent regions A, B, C, and D.
Under the assumption that the two variables are independent, the number of events in region A, here defined as the SR, can be expressed as:
\begin{equation}
    \label{eq:bkg:abcd}
    \NApred = \NB \frac{\NC}{\ND},
\end{equation}
where \NB, \NC, and \ND are the numbers of events in the corresponding regions.
The first variable is chosen to be the \ptmiss, which is the single most discriminating physical observable between signal and background.
The second variable is the output score of a neural network classifier trained to distinguish between signal and background at the event level via the ABCDisCoTEC method, which improves upon the ABCD method in two ways.
The use of a neural network classifier allows for a more powerful separation between signal and background, and the accuracy of the method is directly enforced at training time by measuring the deviation from the ABCD hypothesis, called the nonclosure, as a new loss term:
\begin{equation}
    \label{eq:dnn:loss}
    L_{\text{nonclosure}} = \frac{\abs{\NAa \ND - \NB \NC}}{\NAa \ND + \NB \NC},
\end{equation}
which is required to be below 0.04 when training the event classifier.
This threshold was determined via a hyperparameter scan to optimize the performance with the decorrelation and nonclosure treated as constraints.

The neural network is trained on simulated signal and background events corresponding to the 2018 data-taking conditions that pass the preselection in Section~\ref{sec:EvtReco}, as this is found to generalize well to the full data set.
All background sources are included in the training in proportion to their expected yields after the preselection.
A mixture of 10 different signal models is used for the signal events: starting from the benchmark signal model, seven variations of \mPhi (600, 800, 1000, 1500, 2000, 3000, 4000\GeV) and four of \rinv (0.1, 0.3, 0.5, 0.7).
Equal numbers of signal events and background events are used in each training batch.
Of the signal and background events, 70\% are used for training, while 15\% are reserved for testing, and another 15\% for validation.
The training is performed using the \textsc{PyTorch} package~\cite{Paszke:2019}, and the architecture is a fully connected feedforward neural network with four hidden layers comprising 128, 64, 16, and 8 nodes, respectively, with rectified linear unit activations~\cite{relu} after each but the last layer.
Because the different SVJ production modes outlined in Section~\ref{sec:DataSim} can lead to more than two jets being produced in an event, up to four jets with the highest \pt are used to construct the following input variables:
\begin{itemize}
    \item $z(\widejet_a,\widejet_b)=\min (\pt^a,\pt^b)/(\pt^a+\pt^b)$;
    \item $z(\widejet_a,\ptmiss)=\min (\pt^a,\ptmiss)/(\pt^a+\ptmiss)$;
    \item $\Dphi(\widejet_a,\widejet_b)$;
    \item $\Dphi(\widejet_a,\ptmiss)$;
    \item $\Deta(\widejet_a,\widejet_b)$;
    \item $\dR(\widejet_a,\widejet_b)$;
    \item $\log_{10}{(m_{\widejet_a\widejet_b}/{\GeVns})}$;
    \item $\log_{10}{(\mT(\widejet_a,\ptmiss)/{\GeVns})}$,
\end{itemize}
where $a$ and $b$ are any pair of the included jets in the event, $\pt^a$ and $\pt^b$ are the jet transverse momenta, \Dphi and \Deta are the $\phi$ and $\eta$ separations between the jets, \dR is the angular separation in the $\eta$-$\phi$ plane, $m_{\widejet_a\widejet_b}$ is the invariant mass of the jet pair, and \mT is the transverse mass of the jet and the \ptmiss.
This choice of variables captures the kinematic properties of the multijet system, while naturally accommodating events with less than four jets by setting observables involving a given jet to zero if that jet is not present.
This scheme is inspired by the Lund representation~\cite{Dreyer:2018nbf} applied to event-level variables, treating the \ptmiss as a massless two-vector.

This set of input features is determined primarily by the nonresonant production of the signal events and is largely independent of the parameters of the dark sector that determine the dark showering.
By separating the event into event-level kinematic features and jet-level substructure features, the former can be fully exploited by the supervised event classifier to achieve maximum separation, while the latter are separately targeted by the supervised and unsupervised SVJ taggers to provide results with different tradeoffs between sensitivity and model dependence.

\begin{figure}[!hbtp]
    \centering
    \includegraphics[width=0.4\textwidth]{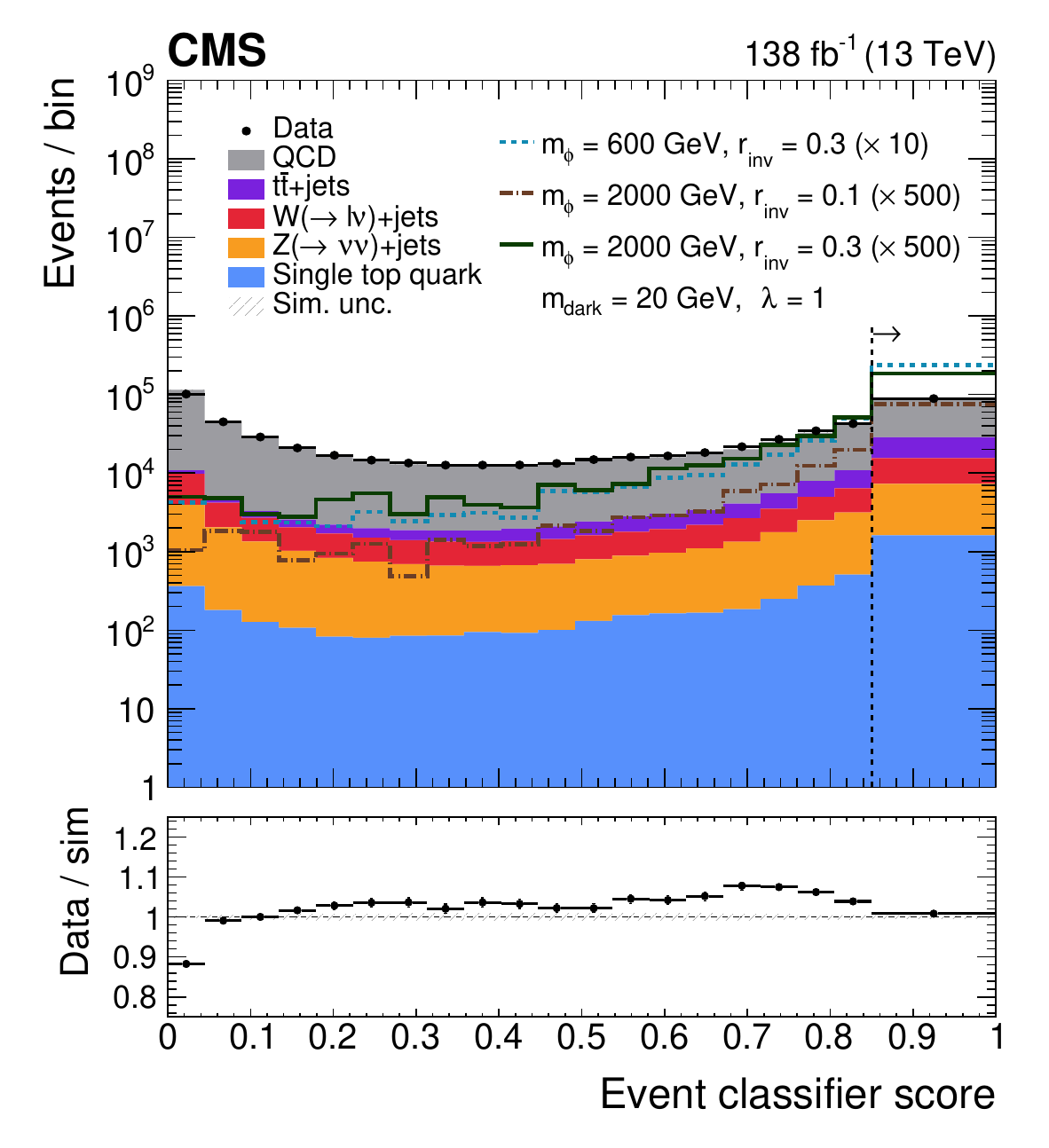}
    \includegraphics[width=0.58\textwidth]{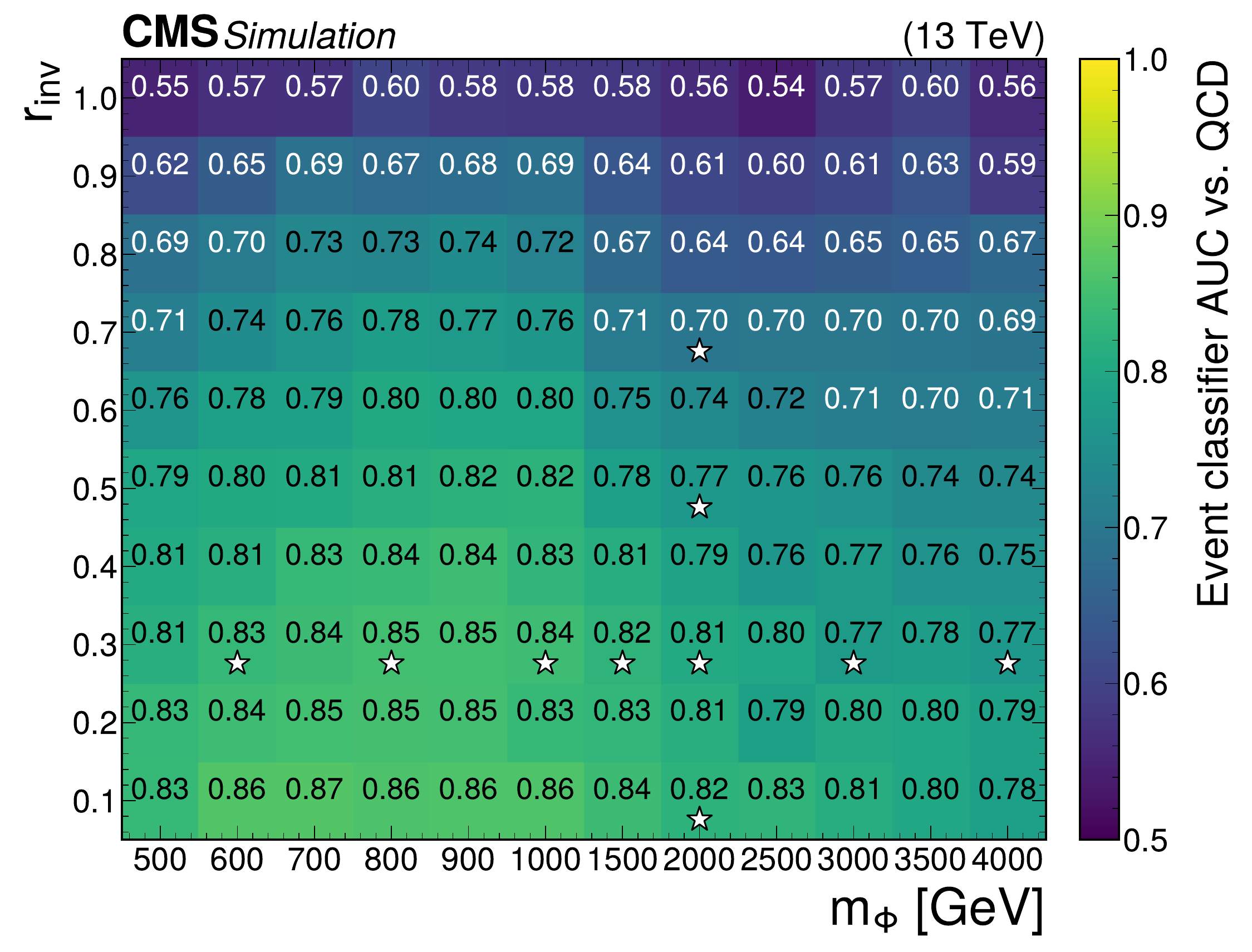}
    \caption{The event classifier output scores for data, background simulation, and a selection of signal models scaled for visibility (left) and the event classifier performance evaluated on a range of signal hypotheses with $\mDark=20\GeV$, $\yukawa=1$, and varying \rinv and \mPhi, against the total background (right), after applying the preselection criteria. The lower panel shows the ratio between the data and the simulated background. The signal models used for training the event classifier are marked with a star. The simulated background score distributions are shown for illustration purposes only; they are not used to predict the SM background in the statistical interpretation. The signal jets are corrected for substructure-related mismodeling using their Lund jet plane density, as described in Section~\ref{sec:EvtReco}. The background simulation has no substructure correction applied. The statistical uncertainties in the data points are smaller than the marker size. The statistical uncertainty in the background simulation from the limited number of simulated events is shown, but is generally small.}
    \label{fig:ABCD_classifier}
\end{figure}

The primary loss term for training the classifier is the binary cross-entropy, which ensures good signal vs. background separation.
This is complemented by a second loss term introduced in Eq.~\eqref{eq:dnn:loss} that enforces the closure of the ABCD method itself, such that the background prediction in the SR is as accurate as possible.
And finally, a third loss term is utilized to minimize the distance correlation (DisCo)~\cite{Gabor:2007} between the \ptmiss and the network score, computed on the background training sample only.
This ensures, as required by the ABCD method, that the classifier and the \ptmiss distributions are uncorrelated.
The modified differential method of multipliers~\cite{Platt:1987} is employed for this multiobjective training: the binary cross-entropy is the primary loss term to be minimized, while the closure and DisCo loss terms are constrained to maximum values of 0.04 and 0.001, respectively.
This setup achieves good discrimination between signal and background, while yielding an ABCD plane with uncorrelated classifier and \ptmiss variables, facilitating a reliable background estimation.
The event classifier scores for data and simulation after preselection, as well as its performance in separating signal hypotheses from the total background quantified via the AUC, are presented in Fig.~\ref{fig:ABCD_classifier}.

As discussed in Section~\ref{sec:SVJtaggers}, events are split into different categories based on the number of tagged jets: $\nsvj=0$, 1, 2, and $\geq 3$.
The $\nsvj=0$ category is included as an SR in the \PNET search as a signal-depleted region to provide constraints in the statistical interpretation of the results and in the WNAE search as a region with nonnegligible signal yield because of the tagger's lower efficiency.
To estimate the background, the ABCD plane is constructed using the \ptmiss and the classifier score in each \nsvj category.
The \ptmiss threshold in the ABCD plane is set to 250\GeV; the relatively low threshold retains sensitivity to signal hypotheses with varying \ptmiss spectra.
The threshold on the event classifier score is set to 0.85 in all \nsvjpn and \nsvjwnae categories.
This threshold was found to yield the best performance while retaining good closure.

The SR is separated into multiple bins of \ptmiss.
The bin boundaries are optimized for the signal significance using simulation separately for the \strategies, setting the upper range of each bin such that it contains a minimum of ten background events.
The binning is allowed to vary between the different \nsvj categories.

The results are extracted by performing a simultaneous fit to the \ptmiss distributions in all categories.
The number of background events in the A, B, C, and D regions of each \nsvj category and \ptmiss bin are scaled together following the ABCD relation in Eq.~\eqref{eq:bkg:abcd}.
Potential signal contamination in regions other than A is included via the overall normalization of the signal, shared between all categories and regions.
For the benchmark signal model, the expected contamination is typically less than 1\%, rising to around 10\% for the $\nsvjpn=2$ and $\nsvjpn \geq 3$ categories in the B and C regions.
The category $\nsvjpn \geq 3$ is treated exceptionally because it has so few expected background events in the ABCD regions that the ratio in Eq.~\eqref{eq:bkg:abcd} cannot be reliably calculated.
Instead, the number of events in region C is scaled by the ratio of events in regions B and D ($\NB/\ND$) to calculate the number of events in the SR.
In this case, the quantities \NB and \ND are taken from the $\nsvjpn = 2$ category, after checking that the ratio $\NB/\ND$ in simulation varies by only 1.6\% between the 0, 1, and 2 \nsvjpn categories and three data-taking years.
An additional systematic uncertainty is included to reflect the potential inaccuracy of this extrapolation.

\section{Systematic uncertainties}\label{sec:Syst}

The systematic uncertainties in the signal simulation primarily originate from two distinct sources.
The first category, experimental uncertainties, arises from detector-related effects, measurement limitations, and event reconstruction.
Such uncertainties often stem from corrections applied to account for differences between data and simulation.
The second category, theoretical uncertainties, reflects variations in the modeling of signal events during event generation.
The systematic uncertainties are further classified based on their impact on the signal.
Normalization uncertainties affect only the overall signal yield, whereas shape uncertainties influence both the event yield and the distributions of kinematic variables.
The experimental uncertainties are treated as uncorrelated across different data-taking years, as they depend on varying detector conditions.
In contrast, theoretical uncertainties are assumed to be correlated across years, given that their sources remain consistent.
Additionally, statistical uncertainties arising from the limited number of simulated signal events are incorporated for each bin of the \ptmiss spectrum.
A summary of the impact of each uncertainty on the total signal yield is presented in Table~\ref{tab:syst}.

\begin{table}[!hbtp]
\topcaption{The range of impacts in \% on the signal yield at the preselection level or on the background yield in region A, for each corresponding systematic uncertainty. Values smaller than 0.01\% are rounded to 0\%. The second column shows the values for the baseline signal hypothesis with $\mPhi = 2\TeV$, $\mDark = 20\GeV$, $\rinv = 0.3$, and $\yukawa = 1$. The third column shows the 10--90\% quantile range across all signal models.}
\centering
\cmsTable{
\begin{tabular}{l @{} c c c}
Uncertainty & Benchmark signal  & Signal (range)      & Background \\
\hline
Statistical & 2.9 & 1.3--4.2 & \NA \\
Integrated luminosity & 0.73 & 0.73 & \NA \\
Trigger efficiency & 2.0 & 2.0 & \NA \\
Jet energy corrections & 1.5--2.6 & 0.58--6.6 & \NA \\
Jet energy resolution & 0.03--0.9 & 0.5--4 & \NA \\
Pileup reweighting & 0 & 0--1.3 & \NA \\
Lund correction syst. uncertainty & 32--38 & 15--54 &\NA \\
Lund correction \pt extrapolation & 4.0--6.5 & 2.3--10 & \NA \\
Lund correction statistical & 9--40 & 7--50 & \NA \\
Lund correction distortion & 26--30 & 8--42 & \NA \\
PDF & 0.12--3.7 & 0--8 & \NA \\
Renormalization/factorization scales & 10 & 3--12 & \NA \\
Per-bin nonclosure & \NA & \NA & 2--65 \\
VR nonclosure & \NA & \NA & 5--50 \\
\end{tabular}
}
\label{tab:syst}
\end{table}

The normalization uncertainties include a 0.73\% uncertainty in the integrated luminosity measurement~\cite{CMS:2021xjt,CMS:2026mqq} and a 2\% uncertainty in the trigger efficiency.
The latter is derived from a data CR that matches the selection of the SR, except that it requires the presence of a muon and the \mindphifull and \ST requirements are removed.
This uncertainty accounts both for potential kinematic differences between the CRs and SR and the statistical uncertainty in the muon trigger CR.
The remaining uncertainties are shape uncertainties.
The uncertainties in the jet energy corrections and the jet energy resolution smearing are assessed, as functions of $\eta$ and \pt~\cite{CMS:2016lmd}, by varying the jet energy scale for all jets in the event and recalculating all jet-related variables, including \ptmiss and \ST.
An uncertainty of up to 1.3\% accounts for the modeling of pileup, derived by varying the total inelastic cross section used to reweight simulated events~\cite{Sirunyan:2018nqx}.

The response of the SVJ taggers discussed in Section~\ref{sec:SVJtaggers} is sensitive to potential mismodeling of the jet substructure in simulation.
To account for this, a correction is applied using a modified version of the method in Ref.~\cite{CMS:2025eyd}, described in detail in Section~\ref{sec:EvtReco}.
The uncertainties in this correction arise from the statistical and systematic components of the measurement of the Lund jet plane density from the CR, the extrapolation of the correction to additional subjet \pt bins, and differences between the kinematic phase space of the signal subjets and the simulated \PW boson subjets utilized in the correction.
The resulting uncertainties in acceptance are in the range 0--79\%, depending on the signal model, \nsvj category, and type of uncertainty.

The theoretical uncertainties include variations in the PDFs, and the renormalization and factorization scales~\cite{Kalogeropoulos:2018cke,Catani:2003zt,Cacciari:2003fi}.
These uncertainties are assessed by reweighting the simulated signal events, considering only the acceptance and shape effects.

The background yields are estimated using the ABCD method described in Section~\ref{sec:BkgdEstimation}.
Two sources of uncertainty are considered to cover any residual nonclosure of the prediction.
The first is calculated per \ptmiss bin in each \nsvj category, utilizing the difference between the predicted \ptmiss distribution from the ABCD method and the true distribution from the background simulation.
It is in the range 2--65\% for the \supervised in all bins and categories except for $\nsvjpn \geq 3$ and 5--20\% per bin, per category for the \unsupervised.
For the $\nsvjpn \geq 3$ category, an uncertainty of 100\% is applied to account for the additional extrapolation from $\nsvjpn=2$ in the background estimation procedure.
The second form of nonclosure uncertainty is an overall normalization uncertainty derived by validating the background prediction directly in data through the use of three validation regions (VRs) made up of the signal-depleted B, C, and D regions of the ABCD plane.
The first VR tests the closure of the method near the \ptmiss boundary; the second tests the closure near the event classifier score boundary; and the third tests both simultaneously.
In each VR, four subregions are defined in order to perform a local closure test, following the method detailed in Section 6 of Ref.~\cite{CMS:2025cvw}.
The largest absolute deviation observed from the procedure is assigned as the VR nonclosure uncertainty separately for each year and \nsvj category, with magnitudes ranging 5--50\% for the \supervised and 5--25\% for the \unsupervised.
For both types of nonclosure uncertainties, the maximum values arise at high values of \ptmiss, which are subject to statistical fluctuations because of the limited numbers of observed events.

\section{Interpretation}\label{sec:Int}

\begin{figure}[ht!]
    \centering
    \includegraphics[width=\textwidth]{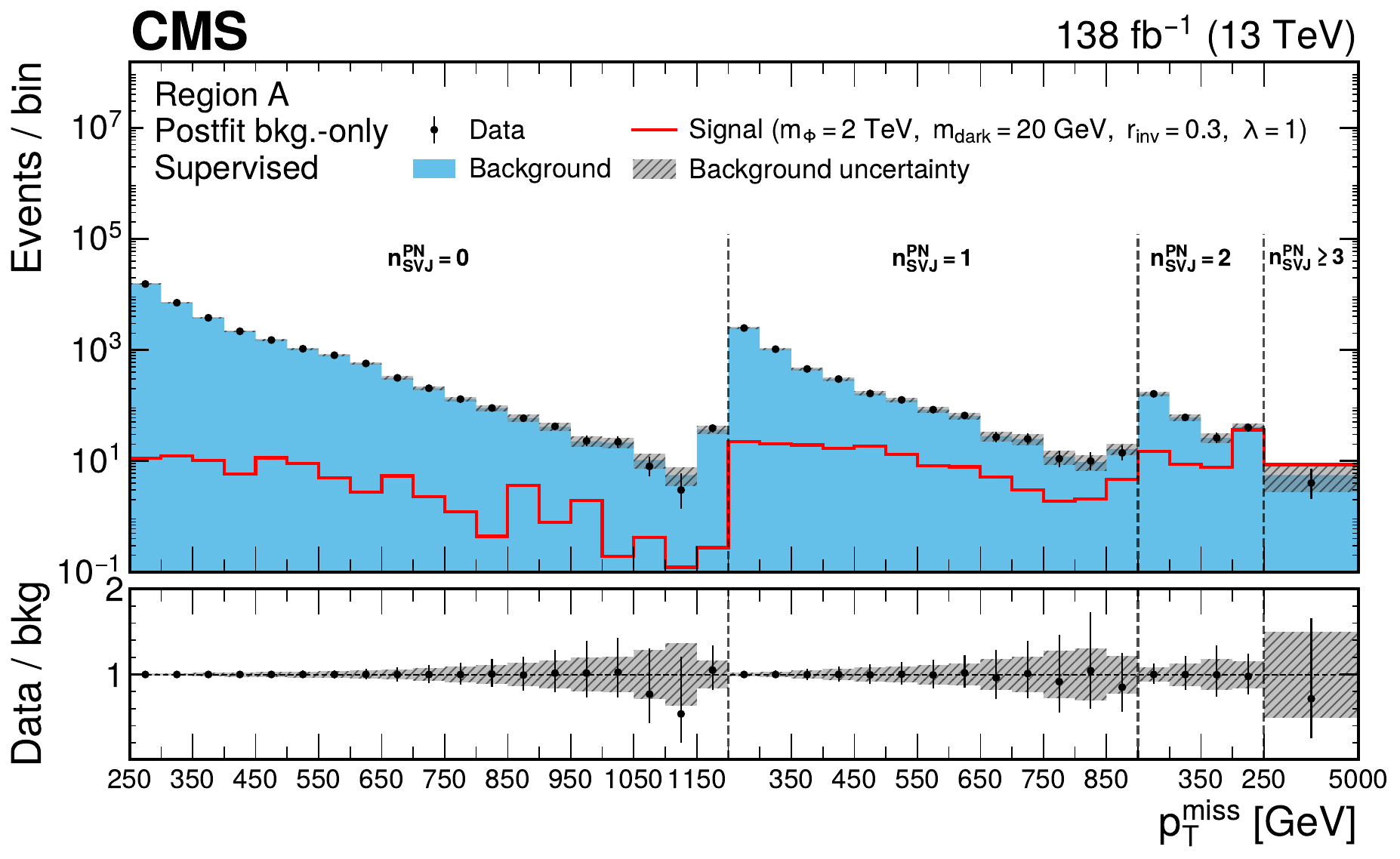}
    \includegraphics[width=\textwidth]{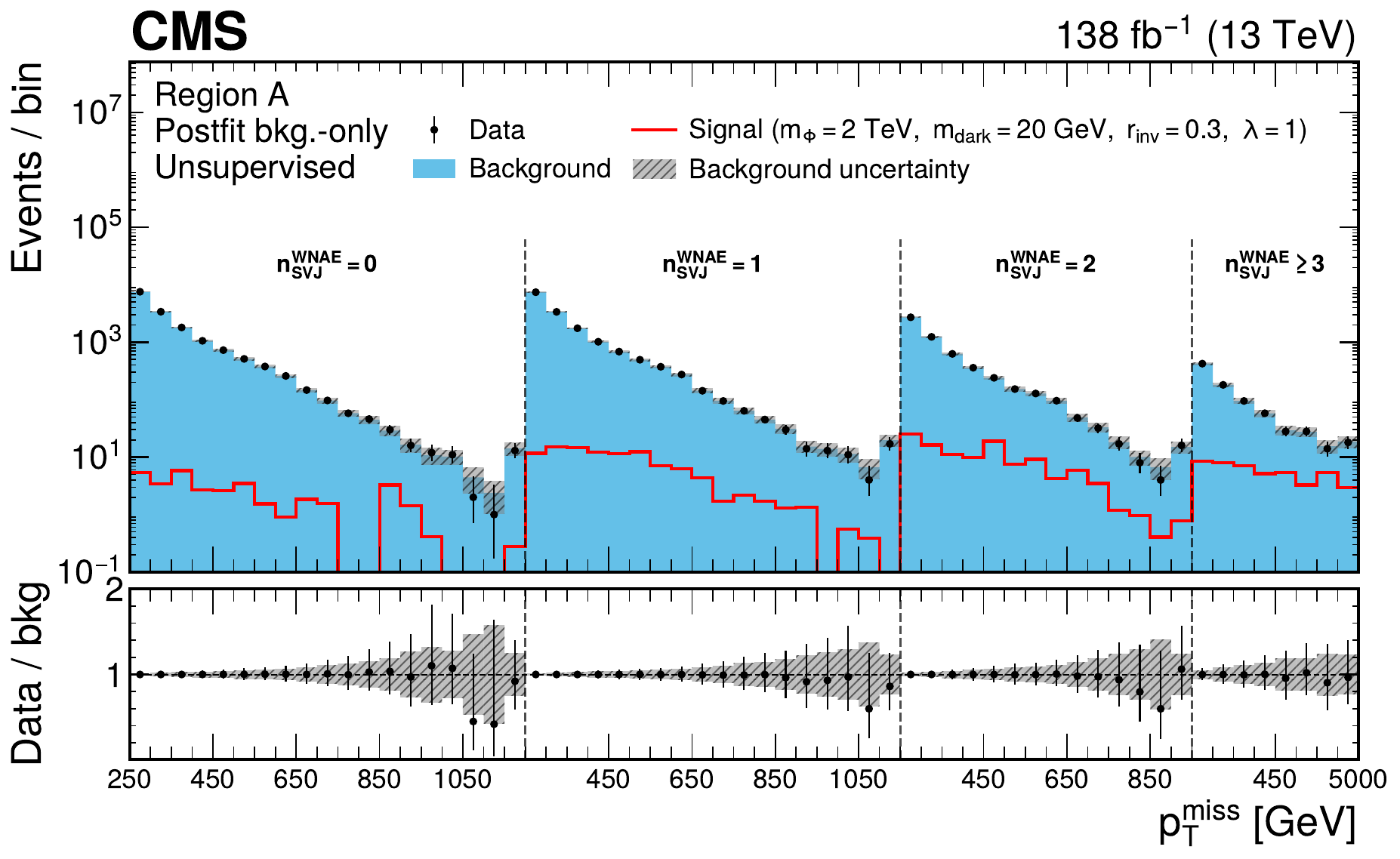}
    \caption{The observed data, the postfit estimated background, and a representative signal model for the \PNET \supervised (upper) and for the WNAE \unsupervised (lower). The background uncertainty includes both systematic and statistical components. The lower panels show the ratio of the observed data to the estimated background. The plots display all three years of data taking combined, while the actual fits consider each year independently.}
    \label{fig:postfit}
\end{figure}

The \strategies are treated separately throughout the analysis, leading to two separate sets of results.
The results are extracted by performing a simultaneous fit to the \ptmiss distributions for all categories for each tagger, using a binned maximum likelihood fit implemented in the framework of the CMS \textsc{Combine} tool~\cite{CMS:2024onh}.
Systematic uncertainties, detailed in Section~\ref{sec:Syst}, are treated as nuisance parameters; the normalization uncertainties are modeled with log-normal priors and the shape uncertainties with Gaussian priors.
In both strategies, the backgrounds are estimated with the ABCDisCoTEC method described in Section~\ref{sec:BkgdEstimation},
with floating parameters for the yield in each \ptmiss bin of the B, C, and D regions used to predict the corresponding A region bin yield.
For each of the \strategies, the different \nsvj categories for each year of data collected are fit separately and statistically combined to achieve better sensitivity.
For each signal hypothesis, the signal strength is treated as a single floating parameter across all regions in the fit.
The final result of the fit, including the estimated background, is shown in Fig.~\ref{fig:postfit}.
The background prediction slightly overestimates the observed data for high \ptmiss values, within uncertainties.
No excess of events is observed with respect to the background prediction.
Limits are placed on the inclusive production cross section \sigmaPhi of the bifundamental mediator \PPhi.

\begin{figure}[ht!]
    \centering
    \includegraphics[width=.49\linewidth]{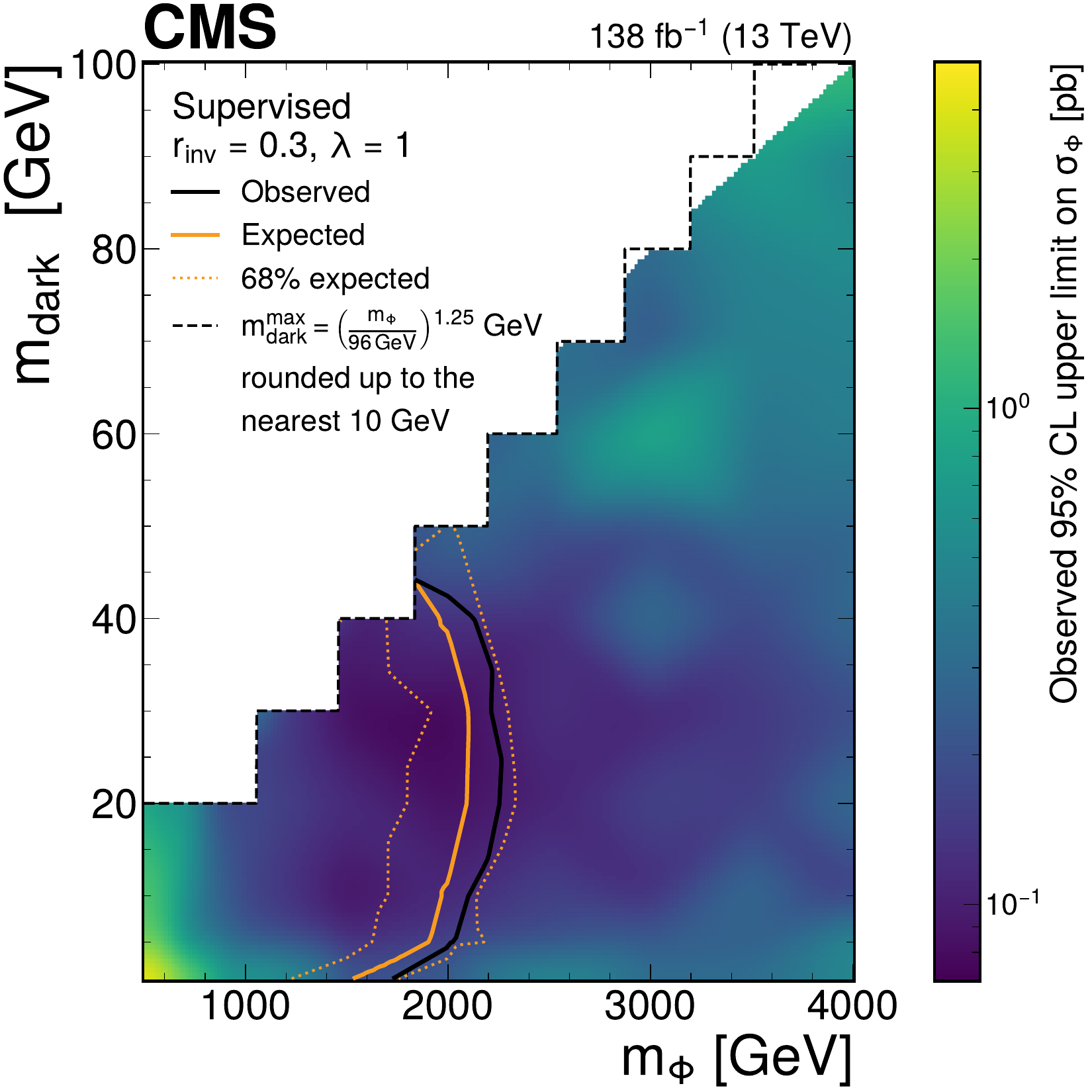}
    \includegraphics[width=.49\linewidth]{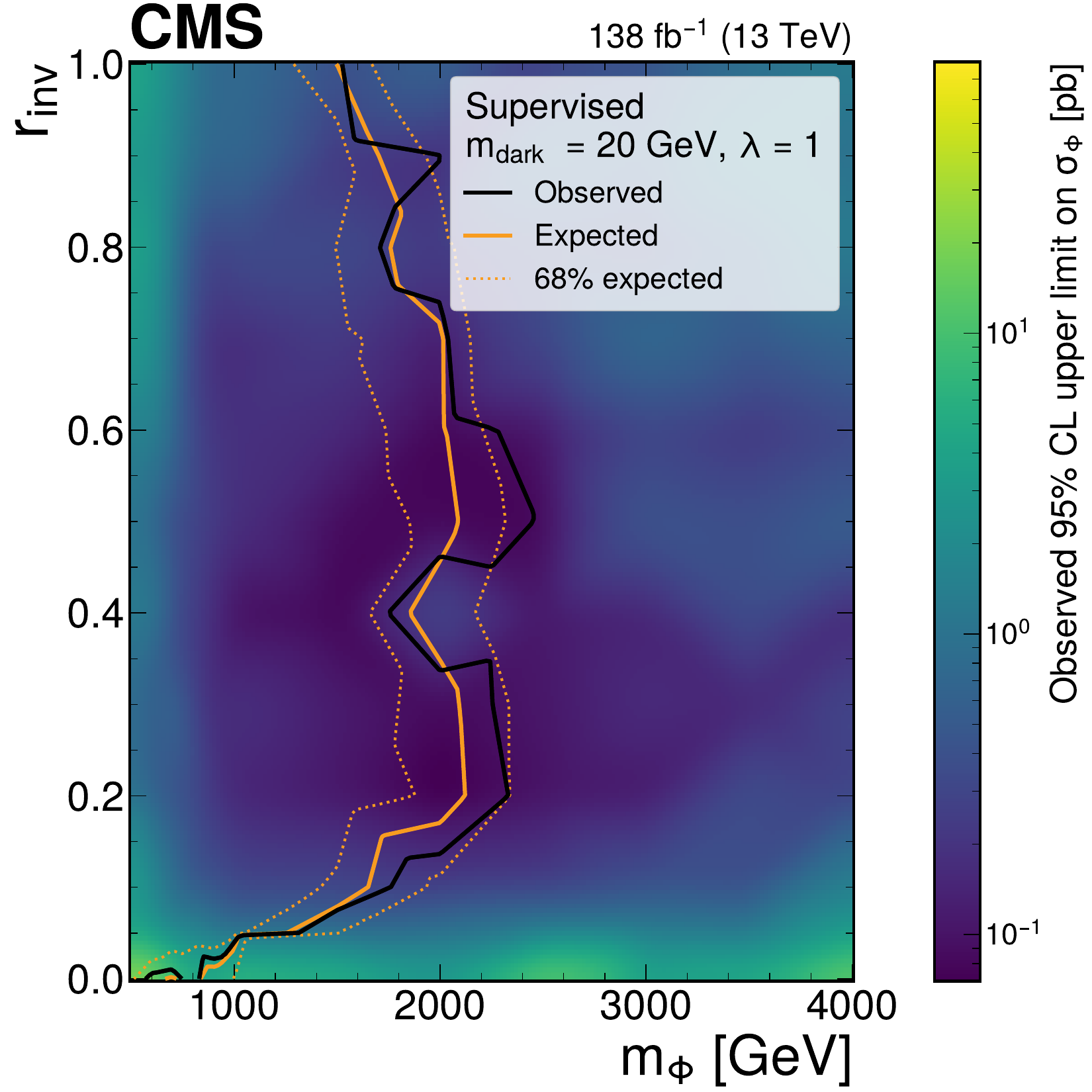}
    \includegraphics[width=.49\linewidth]{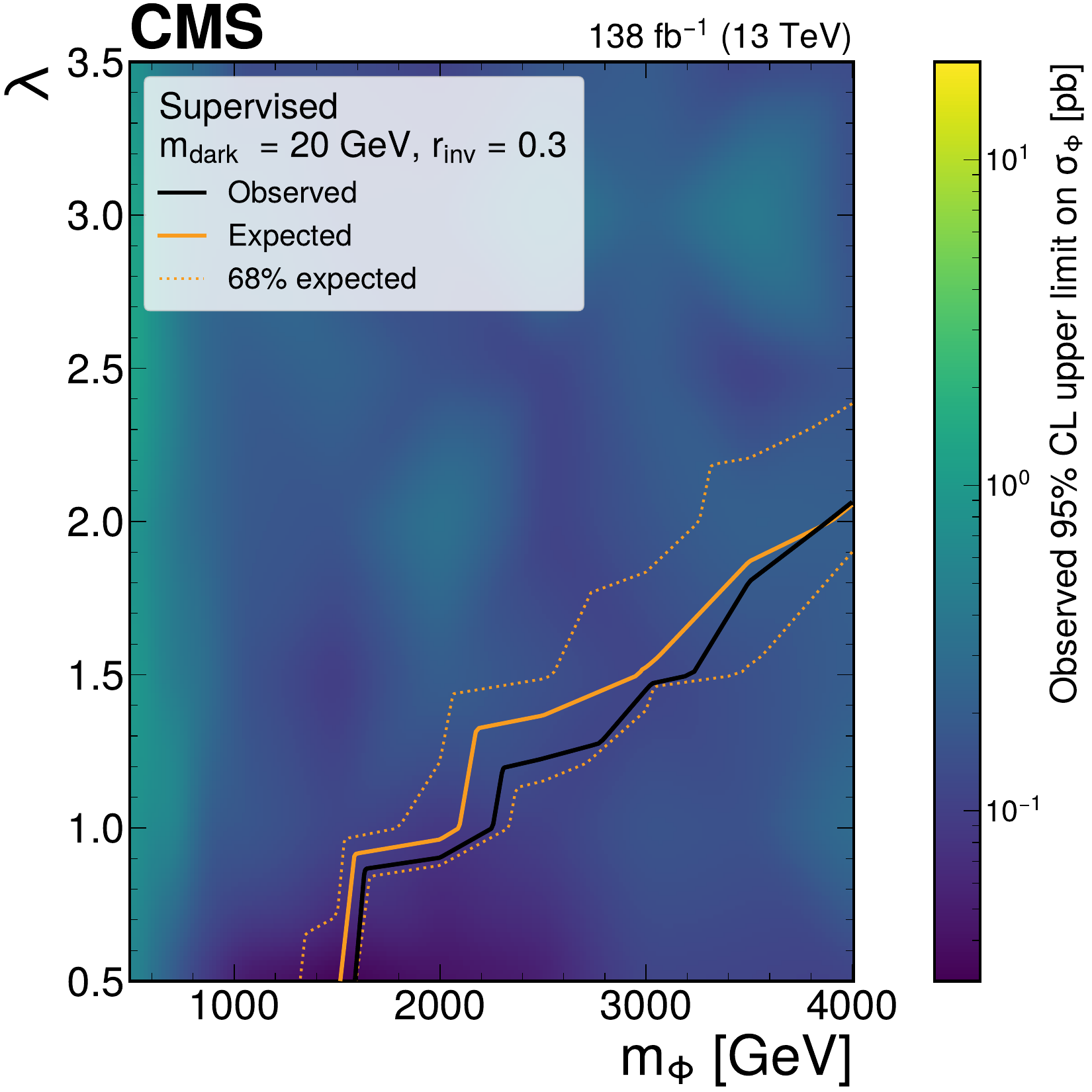}
    \caption{The exclusion limits as a function of \mPhi and \mDark (upper left), \mPhi and \rinv (upper right), and \mPhi and \yukawa (lower), combining all three data-taking years and all \nsvjpn categories, for the \PNET \supervised. The dotted orange lines around the expected limit show the 68\% CL interval. In the upper left plot, the \mDarkMax value is obtained by combining Eqs.~\eqref{eq:lambda_mass} and \eqref{eq:shower_cond} and rounding up to the nearest 10\GeV, with the minimum set to 20\GeV.}
    \label{fig:expected_limits_pn_2D}
\end{figure}

\begin{figure}[ht!]
    \centering
    \includegraphics[width=.49\linewidth]{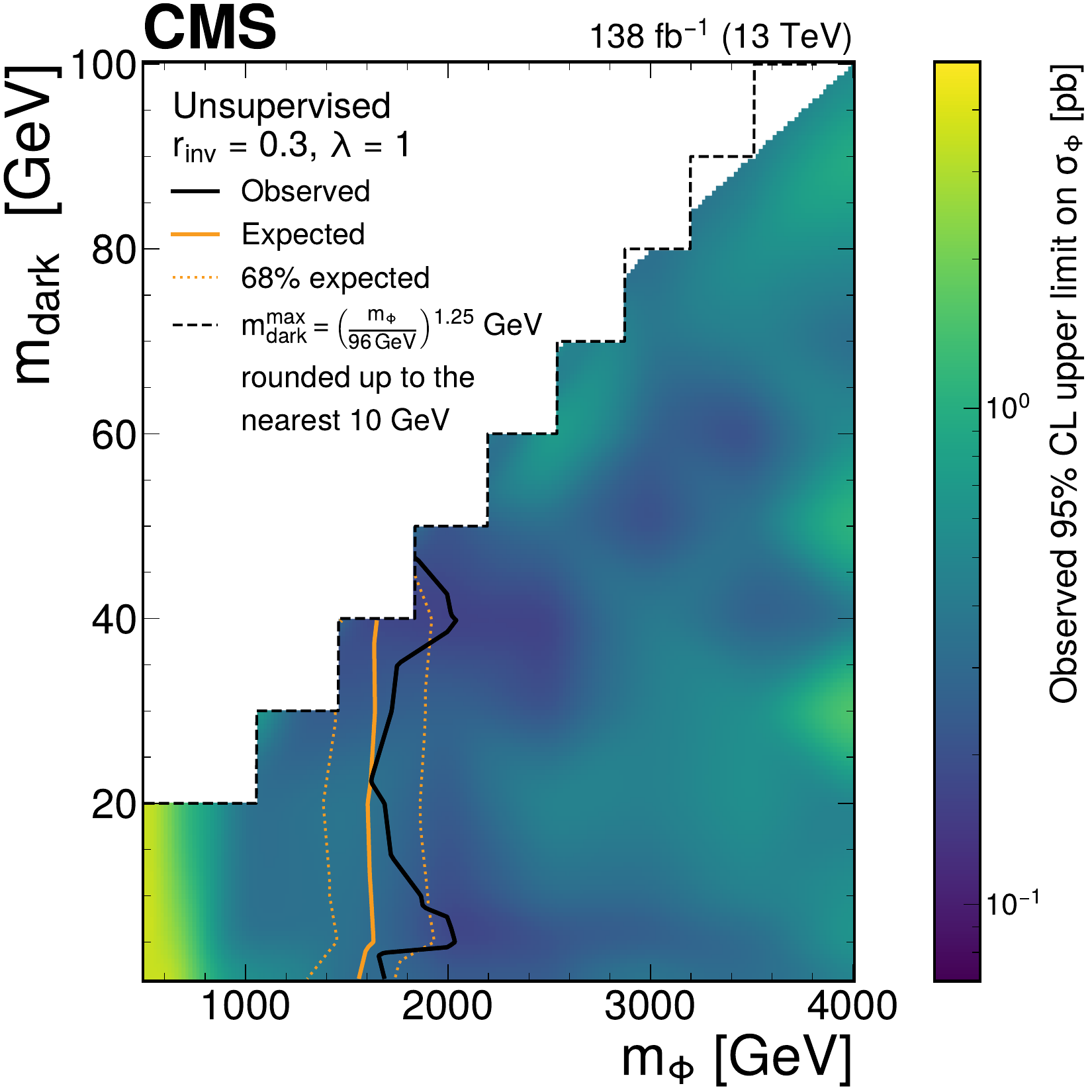}
    \includegraphics[width=.49\linewidth]{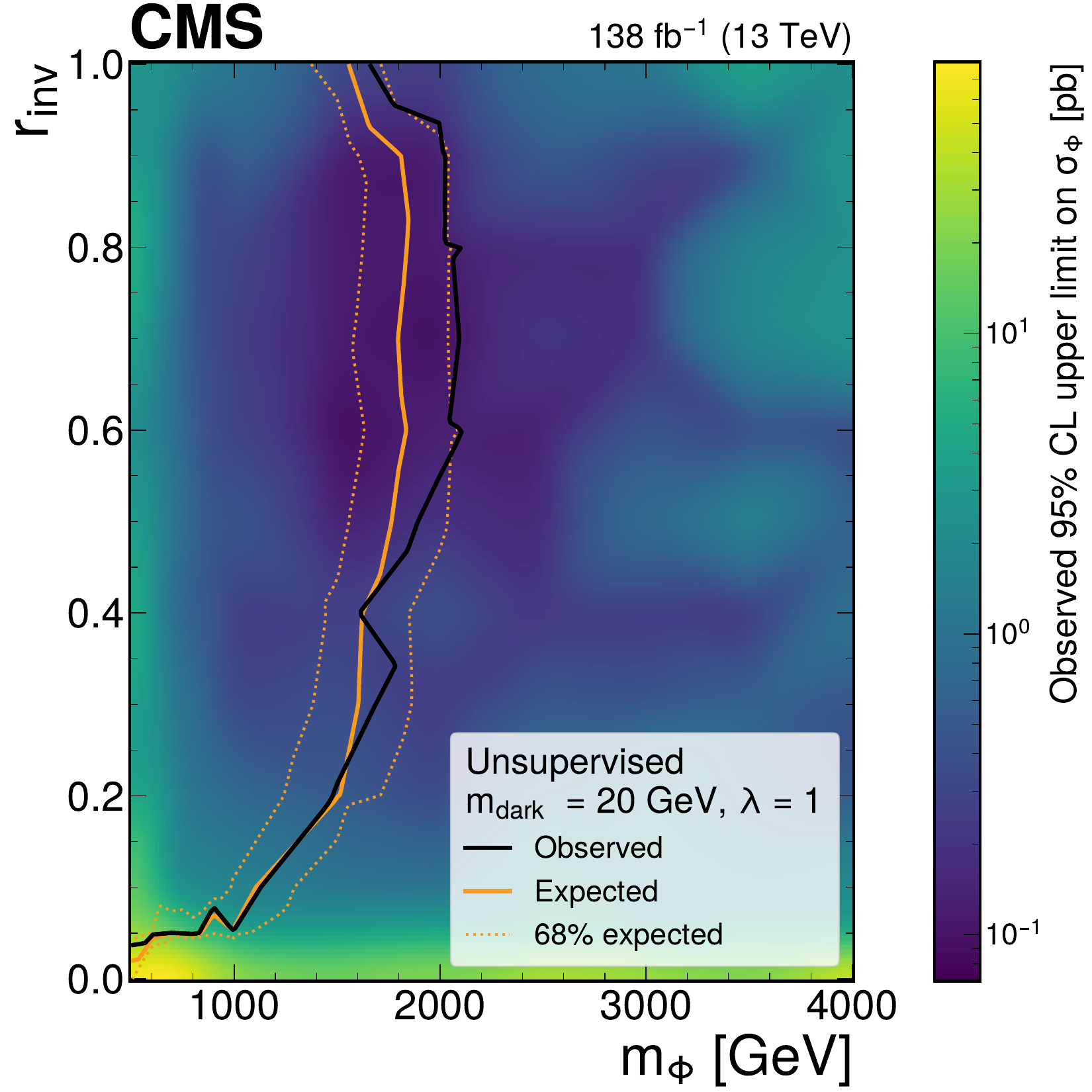}
    \includegraphics[width=.49\linewidth]{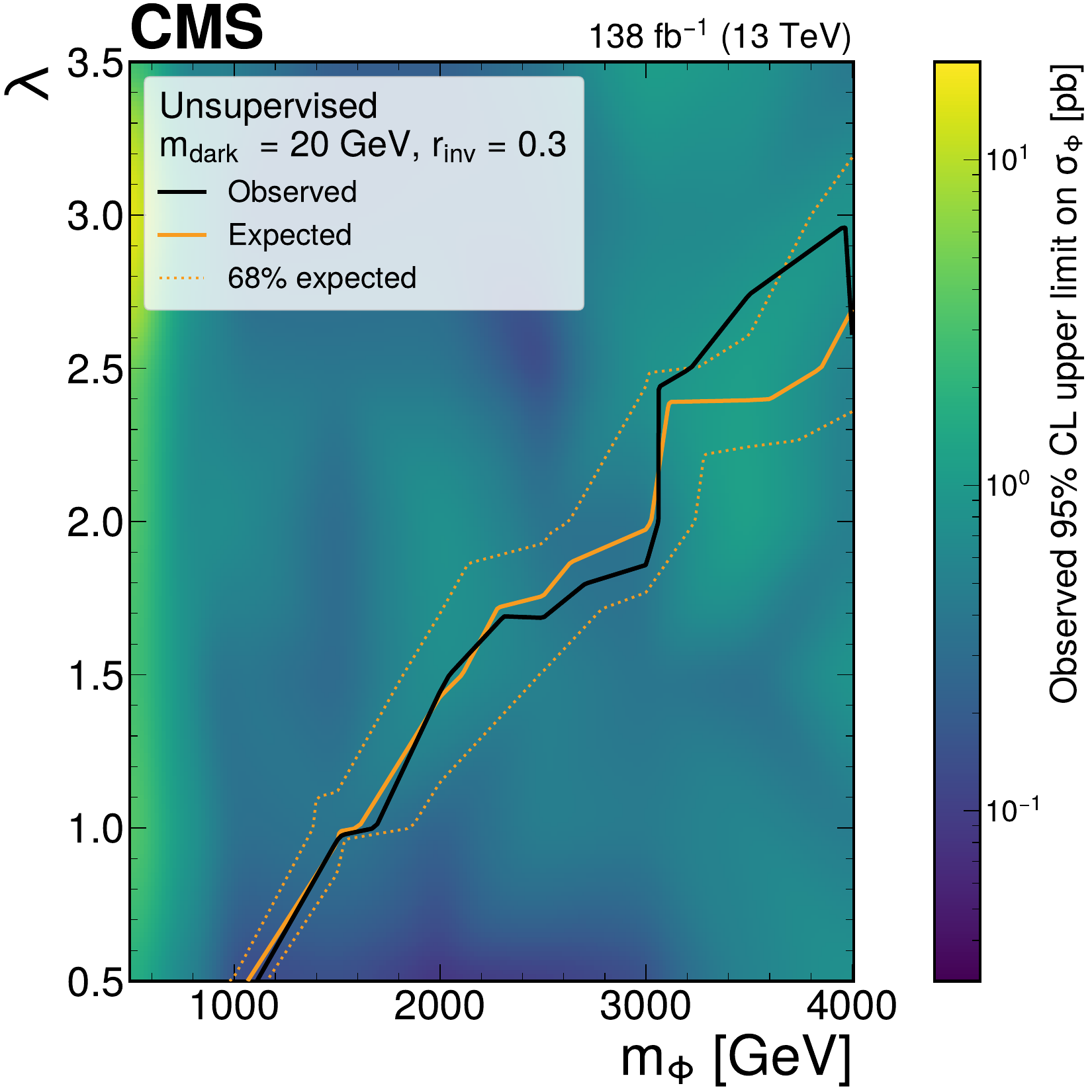}
    \caption{The exclusion limits as a function of \mPhi and \mDark (upper left), \mPhi and \rinv (upper right), and \mPhi and \yukawa (lower), combining all three data-taking years and all \nsvjwnae categories, for the WNAE \unsupervised. The dotted orange lines around the expected limit show the 68\% CL interval. In the upper left plot, the \mDarkMax value is obtained by combining Eqs.~\eqref{eq:lambda_mass} and \eqref{eq:shower_cond} and rounding up to the nearest 10\GeV, with the minimum set to 20\GeV.}
    \label{fig:expected_limits_wnae_2D}
\end{figure}

The cross section \sigmaPhi is constrained at 95\% confidence level (\CL) using the modified frequentist \CLs method~\cite{Junk:1999kv,Read:2002hq} with an asymptotic approximation~\cite{Cowan:2010js}.
The expected and observed limits are presented in the $\mPhi$-$\rinv$, $\mPhi$-$\mDark$, and $\mPhi$-$\yukawa$ planes in Figs.~\ref{fig:expected_limits_pn_2D} and~\ref{fig:expected_limits_wnae_2D} for the \strategies, respectively.
The exclusion contours are determined by comparing the expected and observed limits to the theoretical inclusive cross section for \PPhi-mediated processes, interpolating linearly with Gaussian smoothing applied.

The \supervised excludes mediator masses from 0.5 up to 2.4\TeV for signal models with $\mDark = 20\GeV$ and $\yukawa = 1.0$, with the best performance for signal models with $\rinv = 0.2$ and 0.5.
Its sensitivity decreases as the SVJs become more invisible ($\rinv \to 1$), as there is little visible information available to use for discrimination.
All considered values of \mDark are excluded for mediator masses up to 1.7\TeV, for signal models with $\rinv=0.3$ and $\yukawa = 1.0$.
For mediator masses from 1.7 to 2.0\TeV, all values of \mDark between 5 and 40\GeV are excluded for the same set of signal models.
All values of the dark Yukawa coupling \yukawa are excluded for mediator masses less than 1.6\TeV, and for all mediator masses, $\yukawa > 2$ is excluded for signal models with $\rinv = 0.3$ and $\mDark = 20\GeV$.

The \unsupervised excludes mediator masses from 0.5 up to 2.1\TeV for signal models with $\mDark = 20\GeV$ and $\yukawa = 1.0$, with the best performance for signal models with \rinv values of 0.6--0.9.
For $\rinv = 0.8$, the \unsupervised exceeds the performance of the \supervised, which excludes mediator masses only up to 1.7\TeV, as the supervised tagger was only trained on signal models with \rinv values of 0.3--0.7.
The result of the \unsupervised excludes all considered dark-hadron masses for mediator masses below 1.6\TeV, for signal models with $\rinv = 0.3$ and $\yukawa = 1.0$.
For mediator masses up to 3.0\TeV, \yukawa values above 1.8 are excluded for signal models with $\rinv = 0.3$ and $\mDark = 20\GeV$.
In general, the \supervised is able to exclude a wider range of signal model parameter space for the dark-hadron mass, dark Yukawa coupling, and mediator mass, though the \unsupervised performs better for higher \rinv values.
Both strategies are insensitive near $\rinv = 0$, as those signal models have very low \ptmiss.

The ATLAS search for nonresonant production of SVJs~\cite{ATLAS:2023swa} uses a very different signal model than this search, which hinders a direct comparison.
The ATLAS signal model specifies $\Nfdark=1$ instead of 2 and defines \rinv differently, as the branching fraction for unstable dark hadrons to decay to stable dark hadrons, rather than the overall fraction of stable dark hadrons produced at any stage in the dark shower.
The predicted production cross section for the ATLAS signal model is also higher by roughly an order of magnitude.

\section{Summary}\label{sec:concl}

The first CMS search for nonresonant production of strongly coupled dark matter has been presented.
The search uses data from proton-proton collisions recorded with the CMS detector during the 2016--2018 data-taking period, corresponding to an integrated luminosity of 138\fbinv at a center-of-mass energy of 13\TeV.
The signal model consists of a strongly coupled dark sector with several flavors of dark quarks that shower and hadronize to form stable and unstable bound states called dark hadrons.
The stable dark hadrons are dark-matter candidates, while the unstable dark hadrons decay promptly to standard model (SM) quarks, showering and hadronizing in the visible sector to produce semivisible jets (SVJs) composed of both visible and invisible particles.
This paper focuses on nonresonant production via a massive bifundamental scalar mediator \PPhi with a Yukawa-like coupling between the dark and the SM sectors.
The search probes variations of the mediator mass \mPhi, the dark-hadron mass \mDark, the fraction of stable dark hadrons, and the bifundamental Yukawa coupling \yukawa.

Two complementary strategies are employed to identify SVJs, yielding two sets of results.
The first uses a supervised tagger, \PNET, which is model-dependent and utilizes jet constituents to identify SVJs.
The second uses an unsupervised tagger, the Wasserstein normalized autoencoder, to learn the substructure of SM jets in order to identify SVJs as anomalous jets.
The response of these taggers is sensitive to potential mismodeling of the jet substructure in the signal simulation, which is corrected using a procedure designed for jets with high subjet multiplicity.
This procedure compares the Lund jet plane of the signal subjets, each of which corresponds to one SM quark, to that of subjets from the hadronic decay of a \PW boson in observed data.
For both the \strategies, the SM background is estimated from data using an ``ABCD'' method in which a deep neural network discriminator is trained on event-level kinematic features to be decorrelated from the missing transverse momentum in order to partition events into four regions, separating signal from background.

The \unsupervised is particularly suited to searches for strongly coupled dark sectors, as many details of the signal model can subtly change the resulting jets and their substructure.
Beyond the signal model parameters varied in this search, additional dark-sector parameters such as the number of dark-quark flavors and the number of dark colors, among others, may impact the visible behavior.
The unsupervised tagger is trained independently from any signal model parameter choices and therefore has more consistent sensitivity in order to exclude a broader range of possible variations.

The \supervisedA places 95\% confidence level exclusion limits on the production cross sections for different signal models, which can be interpreted to exclude ranges of signal model parameters.
These exclusions include $0.5 < \mPhi < 2.4\TeV$, depending on the other signal model parameters.
All considered values of \mDark are excluded for mediator masses below 1.7\TeV, depending on other signal parameters.
For mediator masses less than 1.6\TeV, all considered variations of \yukawa are excluded; for the full range of mediator masses considered, $\yukawa > 2$ is excluded.
Similarly, the \unsupervisedA excludes $0.5 < \mPhi < 2.1\TeV$, depending on other signal model parameters.
The \unsupervisedA excludes all considered \mDark masses for mediator masses up to 1.6\TeV.
For the full range of mediator masses considered, $\yukawa > 3$ is excluded.

These results complement existing searches for composite dark matter originating from strongly coupled dark sectors.
A set of new signal hypotheses is excluded by placing limits on the inclusive production cross section for signal processes mediated by a massive bifundamental scalar \PPhi.
The use of machine-learning techniques to exploit both jet- and event-level information increases the sensitivity of the search and enables the exclusion of a wide range of models in the targeted parameter space.

\begin{acknowledgments}
We congratulate our colleagues in the CERN accelerator departments for the excellent performance of the LHC and thank the technical and administrative staffs at CERN and at other CMS institutes for their contributions to the success of the CMS effort. In addition, we gratefully acknowledge the computing centers and personnel of the Worldwide LHC Computing Grid and other centers for delivering so effectively the computing infrastructure essential to our analyses. Finally, we acknowledge the enduring support for the construction and operation of the LHC, the CMS detector, and the supporting computing infrastructure provided by the following funding agencies: SC (Armenia), BMFWF and FWF (Austria); FNRS and FWO (Belgium); CNPq, CAPES, FAPERJ, FAPERGS, and FAPESP (Brazil); MES and BNSF (Bulgaria); CERN; CAS, MoST, and NSFC (China); MINCIENCIAS (Colombia); MSES and CSF (Croatia); RIF (Cyprus); SENESCYT (Ecuador); ERC PRG and PSG, TARISTU24-TK10 and MoER TK202 (Estonia); Academy of Finland, MEC, and HIP (Finland); CEA and CNRS/IN2P3 (France); SRNSF (Georgia); BMFTR, DFG, and HGF (Germany); GSRI (Greece); MATE and NKFIH (Hungary); DAE and DST (India); IPM (Iran); SFI (Ireland); INFN (Italy); MSIT and NRF (Republic of Korea); MES (Latvia); LMTLT (Lithuania); MOE and UM (Malaysia); BUAP, CINVESTAV, CONACYT, LNS, SEP, and UASLP-FAI (Mexico); MOS (Montenegro); MBIE (New Zealand); PAEC (Pakistan); MSHE, NSC, and NAWA (Poland); FCT (Portugal); MESTD (Serbia); MICIU/AEI and PCTI (Spain); MOSTR (Sri Lanka); Swiss Funding Agencies (Switzerland); NSTC (Taiwan); MHESI (Thailand); TUBITAK and TENMAK (T\"{u}rkiye); NASU (Ukraine); STFC (United Kingdom); DOE and NSF (USA).

\begin{sloppypar}
\setlength\emergencystretch{\hsize}
\hyphenation{Rachada-pisek} Individuals have received support from the Marie-Curie program and the European Research Council and Horizon 2020 Grant, contract Nos.\ 675440, 724704, 752730, 758316, 765710, 824093, 101115353, 101002207, 101001205, and COST Action CA16108 (European Union); the Leventis Foundation; the Alfred P.\ Sloan Foundation; the Alexander von Humboldt Foundation; the Science Committee, project no. 22rl-037 (Armenia); the Fonds pour la Formation \`a la Recherche dans l'Industrie et dans l'Agriculture (FRIA) and Fonds voor Wetenschappelijk Onderzoek contract No. 1228724N (Belgium); the Beijing Municipal Science \& Technology Commission, No. Z191100007219010, the Fundamental Research Funds for the Central Universities, the Ministry of Science and Technology of China under Grant No. 2023YFA1605804, the Natural Science Foundation of China under Grant No. 12535004, and USTC Research Funds of the Double First-Class Initiative No.\ YD2030002017 (China); the Ministry of Education, Youth and Sports (MEYS) of the Czech Republic; the Shota Rustaveli National Science Foundation (Georgia); the Deutsche Forschungsgemeinschaft (DFG), among others, under Germany's Excellence Strategy -- EXC 2121 ``Quantum Universe" -- 390833306, and under project number 400140256 - GRK2497; the Hellenic Foundation for Research and Innovation (HFRI), Project Number 2288 (Greece); the Hungarian Academy of Sciences, the New National Excellence Program - \'UNKP, the NKFIH research grants K 131991, K 138136, K 143460, K 143477, K 147557, K 146913, K 146914, K 147048, TKP2021-NKTA-64, 2026-4.1.1-MISSZI\'{O}K-2026-00005 and 2025-1.1.5-NEMZ\_KI-2025-00004, and MATE KKP and KKPCs Research Excellence and Flagship Research Groups grants (Hungary); the Council of Science and Industrial Research, India; ICSC -- National Research Center for High Performance Computing, Big Data and Quantum Computing, FAIR -- Future Artificial Intelligence Research, and CUP I53D23001070006 (Mission 4 Component 1), funded by the NextGenerationEU program, the Italian Ministry of University and Research (MUR) under Bando PRIN 2022 -- CUP I53C24002390006, PRIN PRIMULA 2022RBYK7T (Italy); the Latvian Council of Science; the Ministry of Science and Higher Education, project no. 2022/WK/14, and the National Science Center, contracts Opus 2021/41/B/ST2/01369, 2021/43/B/ST2/01552, 2023/49/B/ST2/03273, and the NAWA contract BPN/PPO/2021/1/00011 (Poland); the Funda\c{c}\~ao para a Ci\^encia e a Tecnologia (Portugal); the National Priorities Research Program by Qatar National Research Fund; MICIU/AEI/10.13039/501100011033, ERDF/EU, ``European Union NextGenerationEU/PRTR", projects PID2022-142604OB-C21, PID2022-139519OB-C21, PID2023-147706NB-I00, PID2023-148896NB-I00, PID2023-146983NB-I00, PID2023-147115NB-I00, PID2023-148418NB-C41, PID2023-148418NB-C42, PID2023-148418NB-C43, PID2023-148418NB-C44, PID2024-158190NB-C22, RYC2021-033305-I, RYC2024-048719-I, CNS2023-144781, CNS2024-154769 and Plan de Ciencia, Tecnolog{\'i}a e Innovaci{\'o}n de Asturias, Spain; the Chulalongkorn Academic into Its 2nd Century Project Advancement Project, the National Science, Research and Innovation Fund program IND\_FF\_68\_369\_2300\_097, and the Program Management Unit for Human Resources \& Institutional Development, Research and Innovation, grant B39G680009 (Thailand); the Eric \& Wendy Schmidt Fund for Strategic Innovation through the CERN Next Generation Triggers project under grant agreement number SIF-2023-004; the Kavli Foundation; the Nvidia Corporation; the SuperMicro Corporation; the Welch Foundation, contract C-1845; and the Weston Havens Foundation (USA).
\end{sloppypar}
\end{acknowledgments}\section*{Data availability} Release and preservation of data used by the CMS Collaboration as the basis for publications is guided by the  \href{https://doi.org/10.7483/OPENDATA.CMS.1BNU.8V1W}{CMS data preservation, re-use and open access policy}.

\bibliography{auto_generated}
\cleardoublepage \section{The CMS Collaboration \label{app:collab}}\begin{sloppypar}\hyphenpenalty=5000\widowpenalty=500\clubpenalty=5000\cmsinstitute{Yerevan Physics Institute, Yerevan, Armenia}
{\tolerance=6000
A.~Gevorgyan\cmsorcid{0000-0003-2751-9489}, A.~Hayrapetyan, V.~Makarenko\cmsorcid{0000-0002-8406-8605}, A.~Tumasyan\cmsAuthorMark{1}\cmsorcid{0009-0000-0684-6742}
\par}
\cmsinstitute{Marietta Blau Institute for Particle Physics, Vienna, Austria}
{\tolerance=6000
W.~Adam\cmsorcid{0000-0001-9099-4341}, L.~Benato\cmsorcid{0000-0001-5135-7489}, T.~Bergauer\cmsorcid{0000-0002-5786-0293}, M.~Dragicevic\cmsorcid{0000-0003-1967-6783}, S.~Gundacker\cmsorcid{0000-0003-2087-3266}, A.K.~Guven\cmsorcid{0009-0004-5670-5138}, M.~Jeitler\cmsAuthorMark{2}\cmsorcid{0000-0002-5141-9560}, N.~Krammer\cmsorcid{0000-0002-0548-0985}, A.~Li\cmsorcid{0000-0002-4547-116X}, D.~Liko\cmsorcid{0000-0002-3380-473X}, M.~Matthewman, A.~Pfeiffer\cmsorcid{0000-0001-5328-448X}, J.~Schieck\cmsAuthorMark{2}\cmsorcid{0000-0002-1058-8093}, R.~Sch\"{o}fbeck\cmsAuthorMark{2}\cmsorcid{0000-0002-2332-8784}, M.~Shooshtari\cmsorcid{0009-0004-8882-4887}, N.~Van~Den~Bossche\cmsorcid{0000-0003-2973-4991}, W.~Waltenberger\cmsorcid{0000-0002-6215-7228}, C.E.~Wulz\cmsAuthorMark{2}\cmsorcid{0000-0001-9226-5812}
% ...
\par}
$^{1}$Also at Yerevan State University, Yerevan, Armenia\\
$^{2}$Also at Technische Universit\"{a}t Wien, Vienna, Austria\\
% ...
\end{sloppypar}
\end{document}